# Navigating and Retrieving Information in Immersive Model-Based Design Reviews: An Exploratory Study

Victor Romero[1], Romain Pinquié[1], Frédéric Noel[1]

[1]Univ. Grenoble Alpes, CNRS, Grenoble INP, G-SCOP, Grenoble, France

## Abstract

Digital engineering uses many models from different perspectives, creating a connected set of digital artefacts across a product's life cycle. Designers seeking a holistic view must navigate numerous models and views, requiring domain-specific software, languages, and representations. This can lead to getting lost in scattered information and the cognitive burden of mentally integrating details across diagrams. To overcome these issues, we developed the virtual environment GraphXplore. GraphXplore enhances perceptual and conceptual integration by linking all relevant visual items from different perspectives into an interactive, layered 3D graph displayed in virtual reality, providing a holistic view of the system. We compared GraphXplore with a conventional on-screen setup using a PowerPoint slide deck with model screenshots viewed on a desktop PC. In an experiment with N=33 volunteers (mainly industrial product design postgraduates and professors), we conducted a baseline usability study focused on fundamental information retrieval tasks for model-based design comprehension, as identifying basic model elements is the fundamental prerequisite in design reviews. Our findings indicate that for simple retrieval tasks, the correctness of answers, completion time, recall score, and perceived confidence are comparable in both environments. However, the virtual environment demonstrated practical advantages, achieving a "good" average System Usability Scale (SUS) score of 73.1 compared to the slide setup's borderline score of 66.4. Furthermore, GraphXplore users reported a lower perceived cognitive workload (mean NASA-TLX score of 43.8) compared to the traditional setup (50.4). Future work will enhance this experiment to yield empirical results across massive industrial datasets, refine GraphXplore, and extend to new design review objectives.

## Graphical Abstract

**Interactive, layered 3D graph in virtual reality *VS.* Static slides**

**Which better supports digital threads comprehension, usability, and mental workload in**

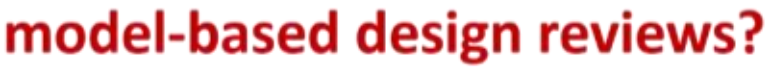

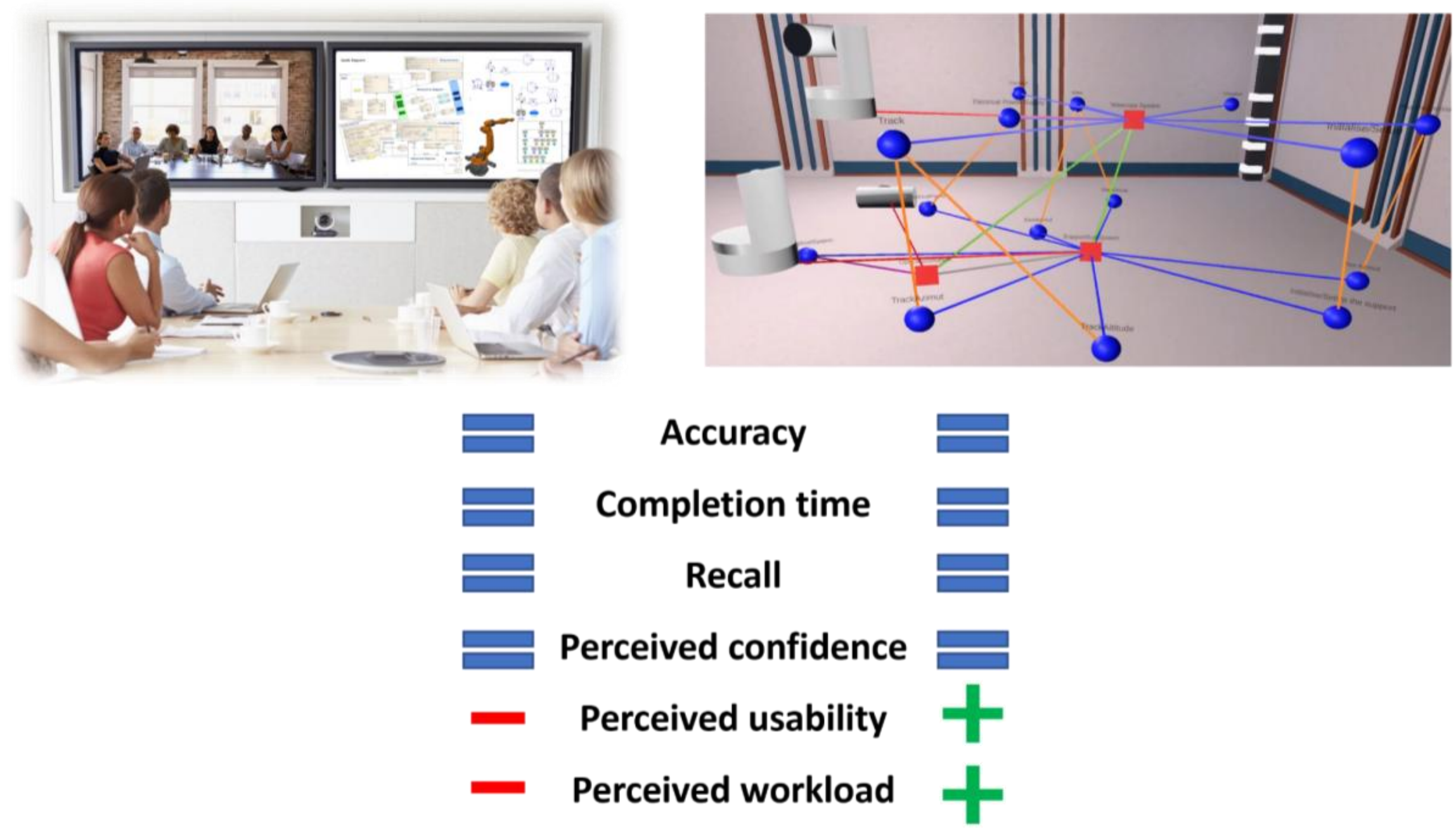

## Highlights

- Introduced GraphXplore, a 3D graph for immersive model-based design reviews

- Improved understanding of esoteric MBSE concepts via 3D spatial interaction
- Compared 3D and slide setups; VR showed better usability and lower mental workload
- Developed a VR environment, benchmark task, and evaluation protocol
- Provided empirical data to advance evidence-based MBSE and VR integration

# 1. Introduction

## 1.1. Research context

The recent shift from Document-Based to Model-Based Systems Engineering (MBSE) brings improvements as well as new challenges. While the benefits of MBSE are primarily backed by perceived rather than measured evidence [1], claimed advantages include establishing a version-controlled single source of truth, standardising notations to reduce ambiguities, and improving data readability and traceability. Nevertheless, MBSE also comes with a set of drawbacks. SysML was heavily influenced by software experts wanting engineered systems to resemble software more closely [2], and several studies point out its limitations [3–5]. It imposes a steep learning curve on non-software engineers, and its existing interfaces often contain undesired software modelling features [4]. that significantly increase deployment and development costs. Finally, existing conceptual systems modelling languages are not well integrated and lack the expressiveness to capture the geometrical aspects of the product (e.g., space allocation and physical interfaces) during the architecture definition phase. Therefore, this paper will address the insufficient visual integration of system views caused by the impact of software engineering on the development of limited MBSE capabilities.

## 1.2.Research problem

With the extensive use of models as the single source of truth for developing and managing systems, when someone wishes to gain a holistic view of the virtual definition of the system and its context, there is no alternative but to navigate through numerous models that require access to and comprehension of domain-specific software and language. Consequently, "*practitioners spend inordinate effort transposing models from rigorous tools to non-structured formats to overcome the acceptance problem. In the process, the connection to the source repository is lost, thereby destroying integrity, reusability, maintainability, and currency of the derived output when the source changes*" [6]. How often have flexible and intuitive tools like PowerPoint slides been preferred for collecting screenshots of system models enriched with notations such as design rationales and traceability links? An alternative is to split the screen into multiple windows or use several screens to visualise various models representing different views simultaneously [7]. Such pragmatic solutions are limited to two or three models due to the restricted size of a PC screen. The challenge of obtaining a bird's eye view of a model-based definition arises not only from the perceptual integration effort to establish interdependence among relevant system elements dispersed across multiple models but also from the conceptual integration effort to generate and refine hypotheses by combining information inferred from the different views [8].

The challenges of accessing and understanding the content of esoteric models and their interdependencies make the design reviews in an MBSE approach difficult. Even the NASA Joint Propulsion Laboratory, a leading organisation in MBSE methods and tools, reports that working-level reviews often use the model directly. However, major gate reviews are still hand-crafted PowerPoint slide decks [9]. Recent studies [10–12] propose new visual analytics tools to facilitate model-based design reviews using digital dashboards that distil important data from MBSE models without providing empirical evidence on usability and utility. To increase the semantic transparency of representations in the manufacturing [13] and construction [14] industries, 3D digital mock-ups reviewed in immersive environments by various stakeholders – including notational non-experts – have replaced codified technical drawings.

By analogy, as new human-model interfaces, we believe that Virtual Reality (VR) has evolved to a stage where users can immerse themselves in a virtual world of data and seamlessly interact with virtual objects. Limited only by its creator's imagination, VR should enhance and ease interactions during model-centric systems design reviews. At the early stages of our research project, these observations motivated us to ask a broad and vague general research question: Does VR enhance and ease the review of a model-centric systems design?

Recent studies [15–18] in software engineering have demonstrated that engineers using VR perform software comprehension tasks in significantly less time while maintaining a comparable level of correctness, especially for complex questions that require examining multiple interconnected visualisations. The developers involved in the distributed software design activities preferred a VR setup to the traditional on-screen one. [17] conclude their paper by suggesting that "*VR may offer advantages in comprehending complex tasks that require navigating* *through multiple interconnected visualisations. However, further experimentation is necessary to validate and reinforce these conclusions.".* Although the application domain is slightly different, this perspective fundamentally aligns with our hypotheses, which assume that an immersive layered 3D graph interactive representation may enable systems engineers to better comprehend linked data in a model-based design review.

The novelty of this proposed representation lies in its fundamental departure from existing VR MBSE applications, which typically just transpose standard 2D SysML diagrams into a 3D virtual environment. Instead of recreating isolated diagrams, our approach leverages a layered 3D property graph to extract and connect cross-domain data, spanning requirements, functions, logical architecture, and physical components, into a single, unified, and explorable digital thread. By utilising spatial layering (altitude) to physically segregate systemic decomposition levels while maintaining explicit relational edges across the entire ecosystem, this visual metaphor is uniquely designed to offload the cognitive burden of mentally integrating fragmented models. It offers a holistic, interactive 'bird's-eye' view of the system's traceability that traditional 2D dashboards, slide decks, and isolated 3D CAD mock-ups inherently fail to provide.

## 1.3. Research method

Qualitative and quantitative research methods were utilised to develop and test GraphXplore, a virtual environment for exploring digital threads during model-based design reviews. Firstly, a literature review was conducted to learn about existing practices and to search for existing virtual environments, designs of experiments, and available test cases. Subsequently, we defined a minimal ontology of key model-based systems engineering concepts. This ontology was implemented in a graph-oriented Neo4J database. We then created a new immersive layered 3D graph representation built upon the ontology to visually explore the model-based design elements. Finally, we designed and conducted a controlled experiment based on the model-based design of a Go-To telescope as a case study to compare two setups:

- **Setup 1 – Slide (control group):** Screenshots of models stored in a PowerPoint slide deck displayed on a desktop PC with a standard keyboard and mouse for interaction. This setup reflects conventional practices.
- **Setup 2 – 3D Graph (experimental group):** GraphXplore is a layered 3D interactive graph representation displayed in a VR Head-Mounted Display (HMD) along with its controllers. It is integrated with the Neo4j database. This setup depicts a simulation of a likely future scenario we aim to analyse.

## 1.4. Research questions

The generic research question that motivated the experiment, "**Can an interactive, layered 3D graph representation in Virtual Reality enhance the comprehension and ease the challenges of navigating digital threads in model-based design reviews, in contrast to conventional slide-based presentations ?**" can be broken down into six testable research questions. Questions 1 and 2 aim to evaluate the influence of the immersive layered 3D graph on task performance during fundamental information retrieval. By conducting a baseline usability study focused on simple retrieval tasks (i.e., understanding how specific model elements relate to each other), we establish the necessary conditions for performing further activities, such as change impact analysis and coverage checks. While GraphXplore aims to ultimately support complex perceptual and conceptual integration, evaluating basic retrieval and usability serves as a necessary foundational step before introducing highly complex cognitive tasks. Although one might interpret that both questions relate to recall, this is not the case, as users can answer questions while exploring the data. Question 3 aims to assess whether technology and representation influence perceived understanding, testing whether new technology and representation mislead systems engineers. Question 4 focuses on an important cognitive function, short-term memory, as it can affect understanding capabilities due to the perceptual and conceptual integration of linked data. Finally, questions 5 and 6 relate to practical usability by systems engineers.

- **RQ1.** Is the engineer's comprehension of a model-based design higher when interacting with an immersive layered 3D Graph representation than with screenshots of models stored in a slide deck?
- **RQ.2** Is the engineer's comprehension of a model-based design faster when interacting with an immersive layered 3D Graph representation than with screenshots of models stored in a slide deck?
- **RQ3.** Is the engineer's perceived confidence in understanding a model-based design higher when interacting with an immersive layered 3D Graph representation than with screenshots of models stored in a slide deck?
- **RQ4.** Is the engineer's short-term memory higher when interacting with an immersive layered 3D Graph representation than when interacting with screenshots of models stored in a slide deck?
- **RQ5.** Is the immersive layered 3D Graph perceived usability higher than the screenshots of models stored in a slide deck for understanding a model-based design?

- **RQ6.** Is the perceived workload to understand a model-based design lower for the immersive layered 3D graph than for the screenshots of models stored in a slide deck?

It is important to recognise that the proposal and research questions are exploratory because we lack empirical evidence. Additionally, systems engineers frequently rely on representations generated by software editors without considering how these visualisations influence users' performance and cognition. As researchers, it is our responsibility to carry out these evaluations.

## 1.5. Statement of contributions

The primary contribution of this paper is an immersive, layered 3D graph (Fig. 1), introduced as a new human-model interaction for exploring digital threads during model-centric design reviews. When conducting a design review, the project team may have various goals, such as elaborating on solutions, evaluating alternative options, and verifying and validating that a solution meets the requirements and the stakeholders' needs. In this study, we focus on the impact of the 3D Graph and Slide setups on the understanding of model-based design.

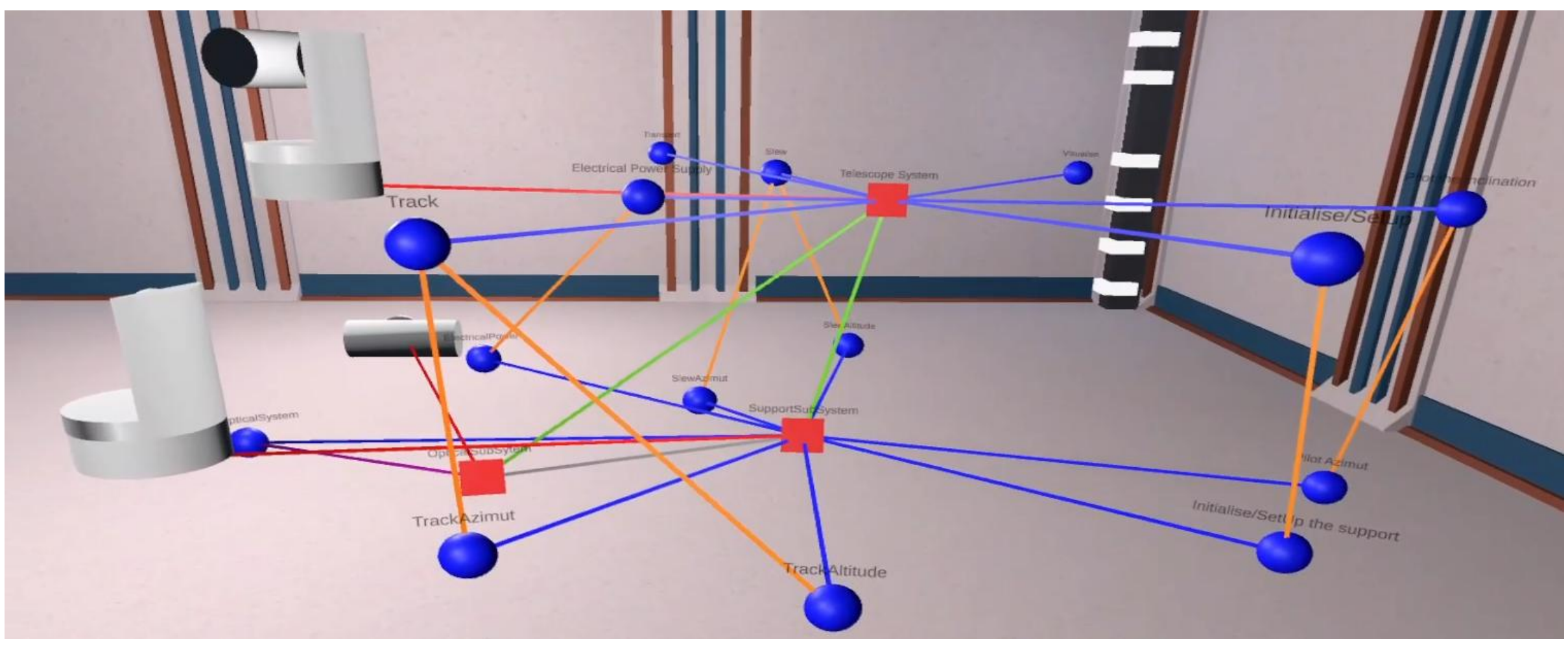

**Fig. 1.** Immersive layered 3D graph visualisation representation to explore digital threads in model-based design (videos are available online[1])

The 3D graph uses 3D spatial positioning within a limitless virtual space to navigate and explore a multi-view integrated design. It also offers new interactions with abstract concepts (e.g., stakeholder, function, requirement, logical unit, geometry) for which previous research [15–20] encourage us to assume that immersive 3D visuals can increase the figurative tangibility, that is, the ability of an actor to mentally grasp (i.e., understand) an abstract concept that can be defined by someone else.

The second contribution is the controlled experiment that compares the 3D Graph setup with the Slide setup. Table 1 presents the results of the hypothesis testing conducted with postgraduate students, professors, and industrial engineering researchers. It is essential to note that, regardless of whether a hypothesis is falsified or corroborated, the significance tests did not enable us to reject the null hypothesis. Although the lack of differentiation suggests the need to define better questions to assess designers' understanding of model-based design, this study stands as an attempt at empirical research in an industry-driven community where anecdotes and case studies provide insufficient evidence to draw conclusions.

**Table 1** Hypotheses and final results

| | | |
|---|---|---|
| ***H1*** | The 3D Graph will be less error-prone than the Slides. | **F** |
| ***H2*** | The completion time is lower with the 3D Graph than with the Slides. | **F** |
| ***H3*** | The perceived confidence is higher with the 3D Graph than with the Slides. | **F** |

[1] https://www.youtube.com/watch?v=eNllFL7QpQE&ab_channel=Vision-R
https://www.youtube.com/watch?v=uiDcBM0KJME&ab_channel=Vision-R
https://www.youtube.com/watch?v=ekarRb1N204&ab_channel=Vision-R

| | | |
|---|---|---|
| ***H4*** | Individuals who use 3D Graph have a better recall score than those who use Slides. | **F** |
| ***H5*** | The perceived usability of the 3D is higher than that of the Slides. | **T** |
| ***H6*** | The perceived workload is lower for the 3D Graph than for the Slides. | **T** |

## 1.6.Limits

This work focuses on understanding a model-based system design where model elements come from different views but are linked together. Although comprehension of a design is a fundamental task for all design review tasks such as ideation, trade studies, validation, verification, or traceability, the results of this study remain limited because such complex design review tasks encompass multiple dimensions (e.g., fluidity of collaboration, mutual understanding, argumentation and reaching consensus, co-operation, etc.).

We also used a single case study, the model-based design of a Go-To telescope. It has a good level of system complexity for conducting controlled experiments without compromising ecological validity, but evidence of the usefulness of VR remains anecdotal. To build confidence in the usefulness of VR beyond this example problem, it is necessary to develop new benchmarks covering the broad variability of engineered systems characteristics (e.g., software systems versus hardware systems with or without embedded software; highly abstract systems like the Future Combat Air System compared to very concrete electromechanical systems; large scale versus small scale systems; etc.).

The independent variable, Setup, also encompasses too many confounding variables (visualisation device, interaction device, 2D/3D representations), which prevents us from explaining the cause-and-effect relationships. The confound could have been avoided by using the same interaction and visualisation devices for both groups and considering the representation as an independent variable. However, conducting an exhaustive set of gold-standard tests to thoroughly examine each plausible hypothesis for each condition would be substantial, so we preferred to conduct an exploratory study.

Ultimately, in addition to the need for a larger dataset whose exploration requires interaction and a large display area, we believe that the questions employed to assess correctness were overly simplistic, which hindered our ability to differentiate effectively among the test conditions. Future experimental designs should aim to create questions that require participants to delve deeply into a large graph to explore the relationships between entities.

## 1.7.Outline

Section 1 has set the context for MBSE reviews. Section 2 reviews the literature on design reviews, emphasising model-based design reviews with and without VR. Section 3 details the design and implementation of GraphXplore, our immersive layered 3D graph. Section 4 presents a controlled experiment that compares the GraphXplore setup with the Slides setup. Sections 5 and 6 report and discuss the results and limitations of the experiment. We conclude the paper in Section 7.

# 2. Literature review

In this literature review, we introduce the concept of digital threads prior to discussing design review practices and tools within document-based and model-based design approaches. Finally, we conclude with an overview of studies examining the use of VR in reviewing software and systems engineering designs.

## 2.1.Digital threads

A central pillar of digital engineering encompassing MBSE is the concept of the digital thread. A digital thread is a linked set of digital artefacts that represent a product’s life cycle, from design to manufacturing, maintenance, and beyond, providing a seamless flow of data that connects all life cycle aspects and whose consistency is actively managed throughout the system's life cycle. Digital threads aim to deploy an integration framework that links heterogeneous information systems and facilitates continuous and synchronised data flow across the various domains of a system's life cycle in dynamic ways, without requiring one-to-one data mapping [21]. With continuous and synchronised data flow across disciplines, digital threads will increase interdepartmental collaboration, streamline product development to reduce production time, and ensure regulatory compliance. A crucial digital engineering activity is developing, using, and maintaining digital threads.

Digital threads, often referred to as big data, can be leveraged using AI or visual analytics, whether in immersive or non-immersive settings. A key technology for implementing digital threads is graph-oriented NoSQL databases such as Neo4j[2] [22–26]. With Neo4j, one can also visualise a flat graph representation with minimal interactions.

## 2.2.Design reviews

Although design reviews are crucial in any design process, the topic is still under-researched [27]. We do not clearly understand what happens during meetings, and the existing tools supporting them (e.g., text editors, voice recordings, whiteboards, visual dashboards) have not disrupted them significantly. Thirty-five years ago, design reviews were already defined as an interaction between designers and reviewers in small meetings involving two to four people, whose purpose was to identify errors (i.e., inconsistencies, inefficiencies, ambiguities, or inflexibilities) in the design [28]. At a time when the design data were captured in documents with very few digital applications, the same authors report that reviewers were already swamped with information. Design reviews evolved from small meetings in the 1990s to become critical milestones in the new product development (NPD) process, utilising the Phase Review Process [29]. Design reviews remained meetings, but they took place at the end of each phase of a stage-gate process with predetermined procedures, organisation, inputs, and outputs. The Phase Review Process introduced critical design reviews, including the Preliminary Design Review (PDR), Critical Design Review (CDR), Initial Design Review (IDR), Final Design Review (FDR), and Test Readiness Review (TRR) [30]. These formal meetings synchronise information about the design and collaboratively evaluate the progress. The review points focus on managing risk, prioritising projects, and allocating resources, with the review team typically being cross-functional, including senior managers from marketing, finance, R&D, and manufacturing [27]. A qualitative analysis of design reviews [31] reveals the predominance of interface negotiation scenarios (e.g., justifications and information requests), with sharing information about the design being more essential than decision-making, exploring, and evaluating. The same authors observed that the goal of a design review varies according to the position in the product development process and reported that process information is exchanged during the early stages, before shifting to product information exchange. In 2019, before the COVID-19 crisis, which undoubtedly impacted our way of cooperating during design reviews, a survey of design reviews reported that a more significant proportion of design reviews had moved from live, co-located meetings to mixed, online, remote, or asynchronous meetings [32]. The taxonomy, including synchronous and co-located (e.g. face-to-face meetings); synchronous and remote (e.g. videoconference and web-sharing); asynchronous and co-located (e.g. routine design activity of a team in the same company); and asynchronous and remote (e.g. design activity of a team involving multiple companies or locations) was borrowed from [33], but the Computer-Supported Cooperative Work matrix was proposed in 1988 by [34]. The same authors [32] also found that the goals of design reviews change throughout the new product development process. Effective communication in the design process is crucial for sharing ideas among individuals with diverse skills and interests, and the shift to a fully digital design process, where physical products are only created after most design work is finished, introduces additional communication challenges, particularly for those outside the engineering field [35].

Among the tasks to be performed during a model-based design review, we will propose a new visual representation to explore the digital threads of a model-based system design. This new human-computer interaction will concentrate on a specific set of needs:

- The solution shall enable design review participants to share a common understanding of model elements.
- The solution shall enable design review participants to trace model elements vertically, from the system level to the subsystems in a top-down and bottom-up direction.
- The solution shall enable design review participants to trace model elements horizontally, that is, throughout the product development phases (e.g., stakeholders, functions, requirements, architecture, geometry definitions) and, more generally, throughout the system life cycle phases.

Although design reviews are poorly understood, we may still question how the shift from a document-centric to a model-based systems engineering approach impacts them.

## 2.3.Design reviews in model-based design

The Joint Propulsion Laboratory (JPL) of NASA, which is a leading actor in the definition of MBSE methods and tools, launched the Integrated Model Centric Engineering (IMCE) Initiative in 2009 [9]. The IMCE initiative identified five MBSE challenges. For each challenge, a desirable “to be” state was defined in 2012, and an actual state based on the widespread adoption of MBSE in the Europa Clipper Mission was reported in 2017. In challenge 1, "Unmanaged Mission Complexity", the author wrote a particular issue: "*Design description comes together only infrequently when preparing for major reviews.*". As a target "to be" state, in 2012, JPL aimed to conduct model-based design reviews that would consist mainly of model inspection and validation. However, the actual state of the Europa Clipper Mission in 2017 reports that informal working-level reviews often use models directly, whereas the "*major gate reviews were still hand-crafted PowerPoint slide decks that tell a story*". Nevertheless, the slide decks contained much more content from the system models. More recently, in 2022, the IRT Saint Exupéry research centre proposed the software EasyMOD [10, 11], which offers capabilities to extract and integrate MBSE data within a review procedure for Review Panel Members at the European Space Agency's Concurrent Design Facility. In a broader sense, EasyMOD tackles four MBSE design review challenges. First, to avoid overwhelming a given review with model and domain-specific information, EasyMOD extracts model views that contain only what is strictly necessary. One crucial point related to this challenge is the necessity to present the extracted information in visuals that do not harm acceptance from the non-specialist audience as it occurs with the highly codified MBSE diagrams [36–39]. Second, EasyMOD links the extracted model views with review objectives and integrates them within a review procedure. Third, EasyMOD enables participants to concurrently create comments on the review procedure or the model views. Finally, EasyMOD facilitates navigation between concepts and views without losing the context of review activities. Although design reviews play a crucial role in the industry, a relatively small body of literature concerns MBSE design reviews. A survey aimed at uncovering the practices and challenges of MBSE reviews (Pinquié et al., 2022) found that participants are often involved in one- to two-hour synchronous, co-located, or remote reviews, where sharing architecture and requirements models is a priority during the early stages of the development process. According to the survey, the main goals of MBSE reviews include requirements validation, establishing a common ground from different views on the system of interest, verifying a design against requirements, and analysing the impacts of a change. During design reviews, meeting participants typically share their screens to display the model in its native software or embed screenshots and videos of the models into a PowerPoint slide deck. The sharing of source models is more of a practice among MBSE experts who have access to the data and software and know how to utilise them. In contrast, embedding models, screenshots, and videos in slides is a common practice among both MBSE experts and non-experts. In terms of difficulties when sharing models during MBSE design reviews, experts and non-experts struggle because the interdependencies between models are primarily implicit, the exploration of models requires to have access and to be knowledgeable about numerous domain-specific modelling languages and software, and part of the information and software capabilities is lost when sharing screenshots of models in slides. Regarding the challenge of dealing with interdependencies between models corresponding to a specific view of the system, survey respondents often share a slide containing screenshots that combine different views and comment on the interdependencies orally. Fewer respondents prefer to share models in multiple windows and comment orally. From these observations, we conclude that MBSE practitioners need new human-model interactions to explore digital threads, especially during MBSE design reviews.

[2] https://neo4j.com/

The literature review reveals a limited understanding of design reviews within an MBSE approach. Despite its relevance for industrialists, this topic attracted very little attention from the scholarly community. Nevertheless, recent collaborative studies between academics and MBSE practitioners have introduced new desktop-based computational tools to support model-based design reviews. Among the technologies, simulation tools demonstrate a value for the early validation and verification of engineered systems during design reviews with a helicopter test pilot [40–42]. According to the INCOSE 2035 [43] and NASA 2029 [44] MBSE visions, immersive visualisation is one of the future characteristics of computer-supported cooperative work. The next chapter will review the use cases of VR applied to MBSE, especially for design reviews.

## 2.4.VR for reviewing model-based designs

In the 80s, mechanical engineering also faced the challenge of integrating various subject-matter experts and managers during design reviews. Indeed, 3D virtual digital mock-ups have replaced the highly codified technical drawings that served as communication models. Today, immersive technologies enable engineers and target customers to navigate a realistic digital mock-up displayed in 3D stereoscopy [45]. VR in design reviews enhances the detection of design errors compared to CAD software on a PC screen, reduces the risk of excluding certain professional groups from the design review process, and facilitates faster entry into the design review [35]. Another experiment [46] reports a significant improvement in participants' ability to understand the geometry of the model correctly and confidently, as well as to comprehend the implications of a proposed design change. A quasi-experiment concluded that VR did not improve the understanding of design representation. However, the same study demonstrated that immersion specifically aided students with lower expertise in better grasping certain design aspects, such as rotation-based mechanisms [47]. Regarding team-based mechanical design reviews, an experiment [48] showed that using an HMD did not significantly affect the number of feedback items; however, there was more problem-related or extrinsic feedback, suggesting a complementarity between the desktop screen and the HMD. In another controlled experiment with teams of designers, [49] found that the VR group demonstrated a significantly higher proportion of creation and assembly actions than the desktop group. These results suggest that integrating VR into design reviews could transform the design process, positioning it as a vital tool for the initial design stages or agile approaches, primarily by increasing the emphasis on creation during the iterative refinement phases. For a more comprehensive overview of immersive VR applications for design reviews, we recommend the systematic literature review done by [45].

The use of VR to visualise representations of concrete objects is quite evident due to a natural mapping between virtual metaphors and real objects, especially if we need depth perception. However, encoding abstract concepts with 3D visuals necessitates the invention of new visual metaphors. Research studies propose immersive virtual environments to review software architecture, where abstract concepts (e.g., functions, packages, lines of code, files, folders, cyclomatic complexity) are encoded with similar symbolic representations in 3D stereoscopy [50, 51] or more iconic representations such as city [15, 16], solar system [52], or island [53, 54] metaphors. Although dependent on prior life experiences, knowledge, and cultural associations, the mapping between familiar visual representations from the real world and abstract concepts is not so natural, raising questions about the notion of metaphor. The mere transposition of 2D symbolic representations into a 3D immersive virtual environment has also been applied to system architecture [55], business processes [56–58] and enterprise architecture [59] diagrammatic representations. Research on the use of city metaphors displayed on a standard computer screen and in VR demonstrated a significant improvement in the correctness of solutions to program comprehension tasks involving diagrams in Eclipse [15]. Another study [16] demonstrated that participants using VR performed the software comprehension tasks in much less time while maintaining a comparable level of correctness, especially for complex questions that required examining multiple interconnected visualisations [17]. In their study, [18] also found no significant difference in the efficiency and recall of design information when using VR; the developers involved in the distributed software design activities are more satisfied with collaboration in VR. By analogy, these recent results encourage us to expect similar outcomes in model-based systems engineering.

Previous works have opted to integrate MBSE with VR instead of replacing diagrammatic representations with immersive virtual environments. For instance, SysML parametric diagrams can be combined with an immersive system representation [60]. Behaviour models in SysML have reduced the preparation efforts required to describe use cases in a VR scene [61]. Recently, a larger number of SysML diagrams have been integrated or visualised in VR [62–64] to support more intuitive and dynamic visual representations of models that are easier to understand for non-systems engineers. The application of VR to support MBSE activities extends beyond the modelling aspect; it has also found applications in the validation and verification of engineered systems within a virtuality-reality continuum, based on a human-centric co-simulation of specification and design models [42]. However, while these approaches successfully transpose 2D methodologies into 3D spaces, they remain largely descriptive and isolated. They often map specific SysML diagrams 1:1 into VR without addressing the broader challenge of perceptual and conceptual integration. By failing to connect cross-domain data into a single, explorable digital thread, these frameworks still require engineers to mentally bridge the gaps between disparate models.

Regrettably, a significant issue in design research stems from the lack of empirical studies, resulting in many claims being made without supporting evidence. This results in a gap between theoretical assertions and practical validation, leading to reliance on unverified statements rather than data-driven conclusions. This literature review confirms that numerous studies [55, 58, 61, 62, 64, 65] that aim to use VR for MBSE do not propose a validation method; therefore, forget that new technology is a means, not an end. Indeed, new technology is only a means to effect changes in a phenomenon of interest that corresponds to the outcome the researcher seeks to improve with the latest technology [66]. For example, Oberhauser (2022) asserts that unlimited virtual space facilitates the depiction and visual navigation of large models; however, this conclusion is not based on empirical data. This paper proposes an immersive virtual environment for reviewing model-based design, as briefly outlined in Romero et al. (2021), and compares it with a traditional on-screen setup through an experiment.

# 3. GraphXplore

This section introduces GraphXplore, an immersive, layered 3D graph functioning as a novel human-model interaction to facilitate the review of model-centric designs, particularly in exploring digital threads. Unlike existing immersive design review frameworks that predominantly focus on physical 3D geometric CAD mock-ups, or frameworks that replicate isolated 2D SysML diagrams within a virtual space, GraphXplore is distinctly built upon a federated property graph data model. This fundamental distinction allows the environment to transcend domain-specific boundaries, merging abstract systems engineering concepts (e.g., requirements, functions) with concrete logical and physical architectures into a single, spatially layered topology. Consequently, rather than acting as a virtual display board for disjointed models, GraphXplore operates as a natively integrated, interactive digital thread. The design solution encompasses defining this graph-oriented data model and mapping it to its corresponding immersive visual metaphors.

## 3.1.GraphXplore architecture

Fig. 2 presents the software architecture described in this section. Since we developed a virtual environment for reviewing models rather than editing them, all relevant data from the models must be imported. The virtual environment connects to a Neo4j database running on the same computer to store and manage the digital threads. Before starting a design review session, the user sends Cypher requests to run parsers that extract elements from the domain models and then transfer them to the Neo4j database. This is done through a 2D environment implemented in Unity. The application calls C# scripts that parse the models and create the Cypher queries that automatically fill the Neo4j database.

Two parsing pipelines are implemented. For the SysML model, a streaming XML parser reads Papyrus/Eclipse UML export files in a two-phase process: first, it extracts raw elements (blocks, requirements, functions, stakeholders, parameters, and their relationships) into intermediate structures; second, it resolves cross-references between these elements to produce fully linked domain objects representing the systems engineering model. For the assembly model, a dedicated parser creates objects representing parts, assemblies, features, joints, and parameters along with their mechanical relationships. Both pipelines produce a unified container of nodes and links. These domain objects are then converted into database-ready representations through a dispatch mechanism based on their type labels. Each adapter generates a Cypher creation statement for its corresponding node or relationship. Nodes receive dual Neo4j labels: a model-level label (SysML or Assembly) and a type-specific label, enabling queries scoped to a single domain model or spanning across models. Cross-model traceability is established through dedicated correspondence relationships that link CAD assembly nodes to their counterpart SysML block nodes.

The nodes and edges of the database are displayed in a virtual 3D world. Primitive shapes and simple colours for easy identification represent nodes that encode abstract concepts. These shapes are selected from a collection of 3D objects stored in the Unity engine and displayed when called by C# scripts. The 3D model can be used for nodes representing the parts or assemblies of the CAD model.

At runtime, the graph visualisation layer queries the Neo4j database using Cypher statements to retrieve the relevant nodes and relationships. Each node type is represented by a specific Unity prefab, allowing different 3D geometries and colours to distinguish blocks, functions, requirements, stakeholders, and parameters at a glance. Nodes are initially arranged in a circular layout and then continuously repositioned by a spring-electric force-directed algorithm implemented through Unity's physics engine. Each node carries a spherical collider and a rigidbody with gravity disabled: when two node colliders overlap, a repulsive force proportional to the overlap depth pushes them apart; conversely, nodes connected by an edge are attracted toward each other until they reach a distance slightly exceeding the sum of their radii. This produces a continuously simulating layout where the graph settles into an equilibrium configuration that visually reflects its connectivity structure. Edges are rendered as lines drawn between their two endpoint positions, updated every physics frame to follow node movement. Different relationship types are differentiated by material colour and line width.

For the CAD model, the system loads a 3D prefab and recursively decomposes it into assembly and part components. Each assembly queries the database to retrieve its corresponding record and establishes parent-child relationships in the scene hierarchy. Leaf parts receive a collision mesh for laser pointer interaction. When a user selects a CAD element, the system traverses up the hierarchy to identify the appropriate level of abstraction and spawns a contextual menu offering actions such as displaying the SysML functions linked to that element through cross-model traceability paths, or decomposing the assembly to reveal its sub-elements.

The VR interaction layer is built on SteamVR with Mirror networking to support multi-user collaborative sessions. A central interaction controller manages four distinct action sets: default navigation, pose tracking, menu navigation, and laser pointer interactions. A hand-attached 3D menu is activated by pressing the controller's menu button; the user navigates its options via the touchpad or stick and confirms a selection with a trigger press. Menu options include populating the database, opening the graph visualisation, displaying the CAD model, and activating laser pointers. When laser pointers are enabled on both hands, callback-driven event handling dispatches click, hover-in, and hover-out events to the appropriate manager depending on the target object. A physical in-scene tablet provides additional interface canvases for node inspection: displaying properties, offering colour annotations for marking elements during the review (e.g., red, green, yellow), managing relationship editing workflows, and enabling the creation of new traceability links between selected nodes.

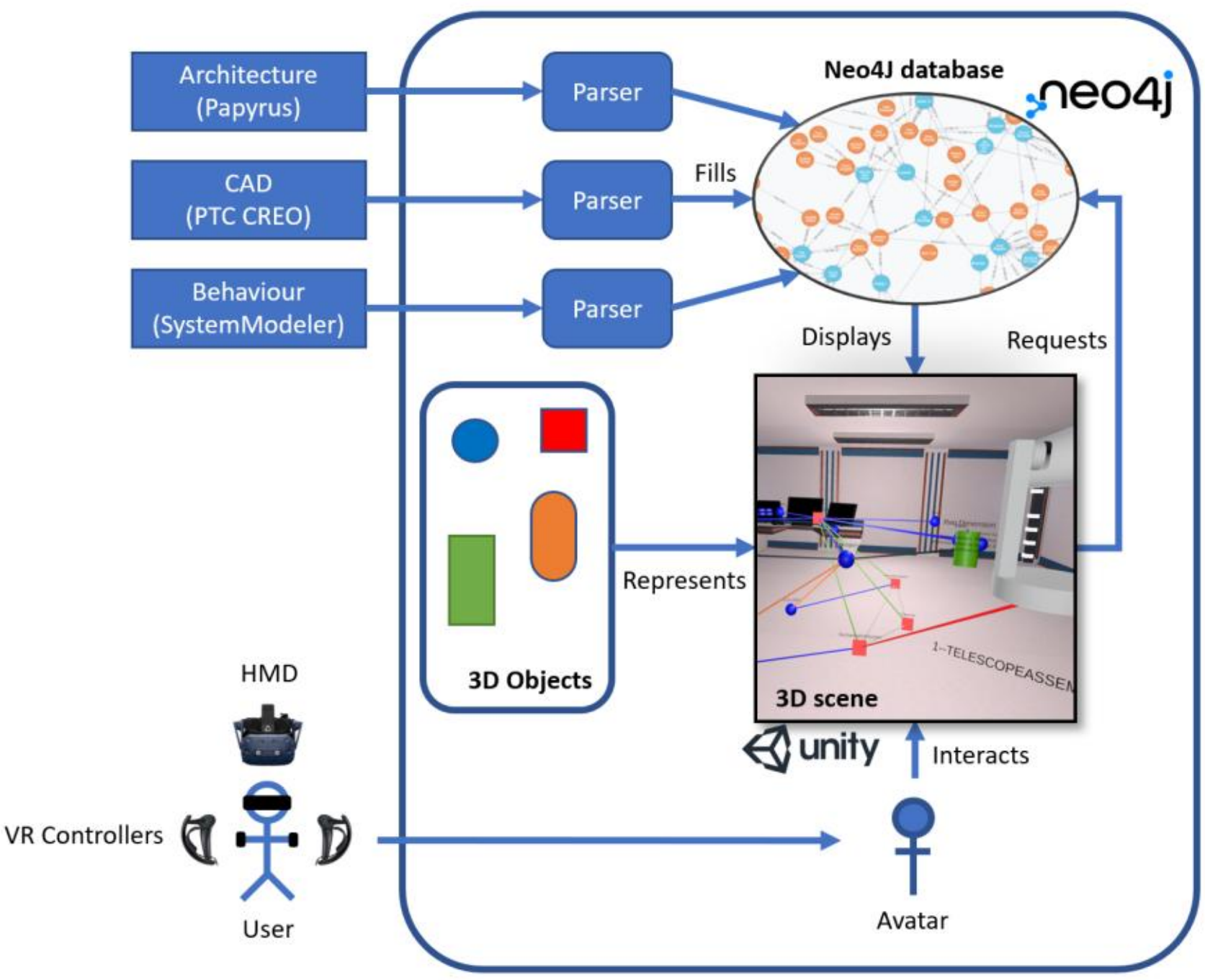

**Fig. 2.** Architecture of GraphXplore

## 3.2.GraphXplore data model

MBSE data are natively stored in MBSE models created using domain-specific languages. Needs and requirements can be retained in requirements management databases or SysML requirement diagrams. Operational scenarios are typically represented in activity and use case diagrams using SysML. The technical functions performed by the system are outlined in architectural diagrams, such as SysML internal block or activity diagrams. CAD models capture preliminary geometric shapes for subsystems, interfaces, and kinematic joints. When early simulation is employed, hybrid discrete and continuous dynamic system models are typically developed using simulation environments such as Modelica or Simulink.

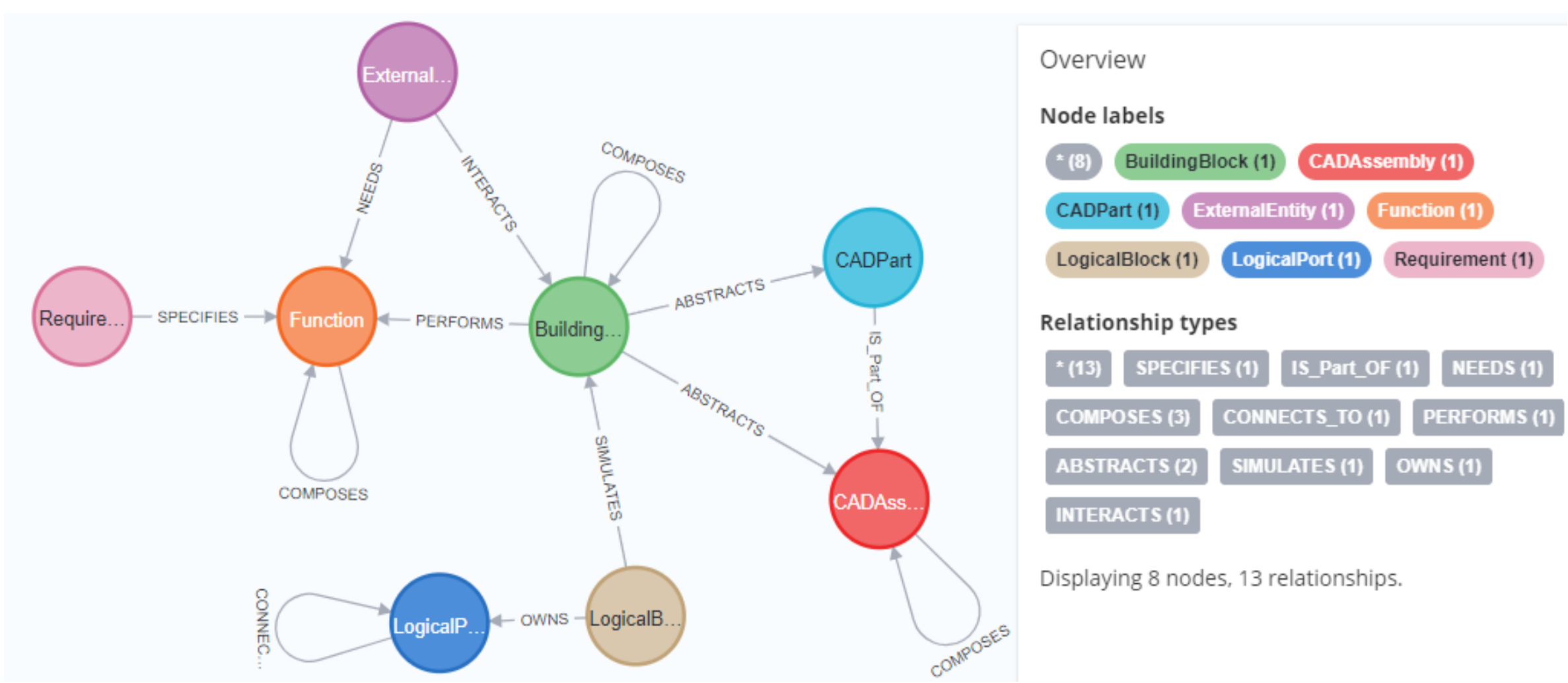

**Fig. 3** Metamodel of the graph-oriented Neo4j database used for the controlled experiment

For several reasons, we propose to store all MBSE data in a federated graph-oriented database for review. Firstly, natural mapping intrinsically links MBSE data (needs, requirements, stakeholders, functions, sub-systems, components, etc.) to a graph data structure. Secondly, a graph is a flexible data structure, which is a crucial feature since each company has its own MBSE method that relies on a unique data model – for instance, Airbus MOFLT [67], Valeo SysCARS [68] or Alstom ASAP [69]. Adding a node or a relationship has minimal impact on the graph-oriented data model compared to SQL data schemas. Consequently, our data model can be easily and quickly tailored to meet any company's practices. Compared to traditional point-to-point API integrations or rigid relational schemas often used to build MBSE digital threads, a graph database provides superior semantic flexibility. Furthermore, while Semantic Web approaches (e.g., OWL/RDF) offer rigorous ontological reasoning for data integration, property graphs like Neo4j are highly optimised for rapid, multi-hop traversals. This query speed is critical for supporting real-time visualisation and interaction in a VR environment without introducing latency. Thirdly, although the volume of data will be small in our case study, the database could contain tens of thousands of elements in an industrial context. Querying relational data in a graph database is more efficient than in any other SQL or NoSQL database. Finally, the graph data structure provides opportunities for graph analytics based on numerous graph algorithms, which are powerful techniques to support MBSE activities such as change impact analysis.

As illustrated in Fig. 3, the data model definition used to prototype the virtual environment comprises eight nodes and fourteen edges, along with their attributes defined in Table 2. The proposed ontology is intentionally kept simple and is derived from an intersection of the main MBSE concepts. It is intended for illustrative purposes and does not aim to encompass all MBSE implementations.

The distinction between building block and logical block concepts may not be immediately evident to readers. A building block is an abstract system resulting from the decomposition of a system of interest from the architectural viewpoint. It is synonymous with the term system element, which is commonly used in systems engineering literature. In SysML, a building block is represented as a block in the Block Definition Diagram or a part property in the Internal Block Diagram. A logical block is a building block viewed from a behavioural perspective. It is a block that models time-dependent behaviour in a system simulation software such as Modelica-based modellers, Simulink, or Amesim. Both should be coherent, as the former defines the entity along with its inputs and outputs, whereas the latter describes its dynamic behaviour.

**Table 2** Attributes of the nodes of the prototype database

|  |  |  |
|---|---|---|
| <External entity> | ID: <String> | Human-readable unique identifier of an external entity |
| | Name: <String> | Name of the external entity |
| <Function> | ID: <String> | Human-readable unique identifier of a function |
| | SystemicLevel: <int> | Systemic level of a function (0 for the system, 1 for the first level of sub-systems, etc.) |
| | Name: <String> | Textual statement of a function |
| <Requirement> | ID: <String> | Human-readable unique identifier of a requirement |
| | SystemicLevel: <int> | Systemic level of a requirement |
| | Name: <String> | Textual statement of a requirement |
| | ReqID: <string> | Identifier of the requirement according to its allocation to a (sub)system (e.g., Req2.3 = requirements three allocated to subsystem 2) |
| | Text: <string> | Textual statement of a requirement |
| <Building block> | ID: <string> | Human-readable unique identifier of a building block |
| | SystemicLevel: <int> | Systemic level of a building block |
| | Name: <string> | Name of a building block |
| <Logical block> | ID: <string> | Human-readable unique identifier of a logical block |
| | Name: <string> | Name of a logical block |
| <Logical port> | ID: <string> | Human-readable unique identifier of a logical port |
| | Name: <string> | Name of a port |
| | Type: <enum> | Type of a port (can be "IN", "OUT" or "IN-OUT") |
| | FlowType: <string> | Type of the flow transiting through a port |
| <CAD Part> | ID: <string> | Human-readable unique identifier of a CAD part |
| | SystemicLevel: <int> | Systemic level of a CAD part |
| | Name: <string> | Name of a CAD part |
| | Volume: <double> | The volume of a CAD part in $m^3$ |
| | Color: <double[]> | Colour of a part encoded in RGB |
| <CAD Assembly> | ID: <string> | Human-readable unique identifier of a CAD assembly |
| | SystemicLevel: <int> | Systemic level of a CAD assembly |
| | Name: <string> | Name of a CAD assembly |

In general, the data modelling aspect is not novel compared to existing MBSE digital thread solutions [21, 24, 25] but aligns with companies' data persistence requirements. The data model is sufficiently simple to remain generic. To address real-world needs, one may explore more detailed aspects, such as ports and flows of logical blocks, kinematic joints between assembly parts, and variants, among others. Adding new nodes and edges will necessitate finding new visuals to encode the corresponding MBSE concepts.

## 3.3.GraphXplore visual metaphors

As discussed in the previous sections, encoding data with a graph provides a flexible, schemaless database solution that effectively facilitates the creation, reading, updating, and deletion of linked data. The graph serves as the data structure and a 3D interactive visual representation. Nodes and edges represent MBSE data and their interdependencies, respectively. An interdependency is an explicit relationship captured within the graph-oriented data model. By navigating through interconnected nodes, the end user reconstructs the traceability of the data.

Compared to a graph in two dimensions, a layered 3D graph offers greater flexibility in positioning the vertices and edges, and crossing can always be avoided [70]. Furthermore, due to the limitations of two-dimensional output media, traditional displays can only provide a restricted resolution and display area. In contrast, VR devices (e.g., HMDs or CAVEs) enable users to visualise the 3D graph in stereoscopic 3D with natural navigational operations (e.g., rotation, shifting, zooming) while retaining their overall mental map. Recent experiments have also shown that, for data analysis, participants need to walk more in 2.5D than in a 3D graph to observe all layers, which increases task completion times [19]. Other studies revealed that compared to 2D, immersive 3D graphs require more time to complete tasks (…), but the interpretation of network structures (e.g., finding the shortest path) is more precise. Participants can effectively manage occlusions in 3D by simply moving their heads to alter the viewpoint [20].

Alongside the visual metaphor of altitude, the graph nodes are presented in a manner that remains consistently readable. The force-directed graph algorithm inspires the script used in the software. In the virtual scene, the nodes regulate their positions by repelling other nodes to prevent excessive proximity while attracting those that share a link, ensuring linked nodes remain sufficiently close to visually represent the related data in a distinct location.

The design of the layered 3D graph relies on guidelines, especially the Physics of Notation [71] and the Cognitive Integration Process [72], which recommend quality criteria for designing and evaluating visual notations:

- **[Semiotic clarity]** The graphical representation of systems engineering concepts is semiotically clear (no synographs and homographs), as there is a 1:1 correspondence between semantic constructs and graphical symbols. It maximises expressiveness (by eliminating symbol deficit), precision (by eliminating symbol overload), and parsimony (by eliminating symbol redundancy and excess).
- **[Perceptual discriminability, Visual expressiveness and Graphic economy]** Perceptual discriminability refers to the ease and accuracy with which graphical symbols can be distinguished. Whereas perceptual discriminability assesses pairwise visual variation among symbols, visual expressiveness evaluates visual variation across the entire visual vocabulary. GraphXplore is perceptually direct as it employs shapes and colours that convey the meaning of its referent concepts. It enhances the differentiation of graphical symbols since each type of node and relationship has its distinct shape and colour. A limit of two visual variables is deliberately set to keep the representation cognitively manageable. Colour is favoured as it serves as a cognitively effective visual variable. Colour differences are detected three times faster than shapes and are more easily remembered. However, colours are sensitive to variations in visual perception and should be used for redundant coding. Therefore, each node type is also visually encoded with a different shape. Redundant coding, which suggests using multiple visual variables (e.g., shapes and colours) to differentiate between them, increases the visual distance between symbols, thus reducing errors and minimising noise.
- **[Semantic transparency]** A visual notation is semantically transparent if the meaning of a symbol can, to some extent, be inferred from its appearance. The graph metaphor of GraphXplore (Fig. 4) provides cues to the symbols' meanings, as form (canonical nodes and line edges) suggests content (entities and relationships). However, it remains semantically conventional since the correspondence between the appearance of symbols and their meanings is purely arbitrary. Indeed, canonical shapes such as spheres, cubes, cylinders, and ellipsoids represent data as nodes. One may prefer to encode concrete concepts with iconic representations. Nevertheless, symbolic visual metaphors are more flexible than iconic ones, as the end-user does not need to create or search for a specific 3D asset that best represents the MBSE concept for encoding. Our decision to employ simple geometric primitives rather than complex, iconic 3D models is also grounded in cognitive load theory and VR design heuristics. Minimising the polygon count and visual clutter of individual nodes prevents perceptual overload, allowing users to focus entirely on the topological structure, the relationships, rather than deciphering intricate 3D geometries. Moreover, most MBSE data (e.g., functions, requirements, architecture blocks, flows, etc.) are abstract concepts without a natural mapping to an iconic representation. To enhance semantic transparency, GraphXplore could incorporate a 3D sign (e.g., F for function, R for requirement, B for block, etc.) within each 3D graphical object of a node type in a future version.

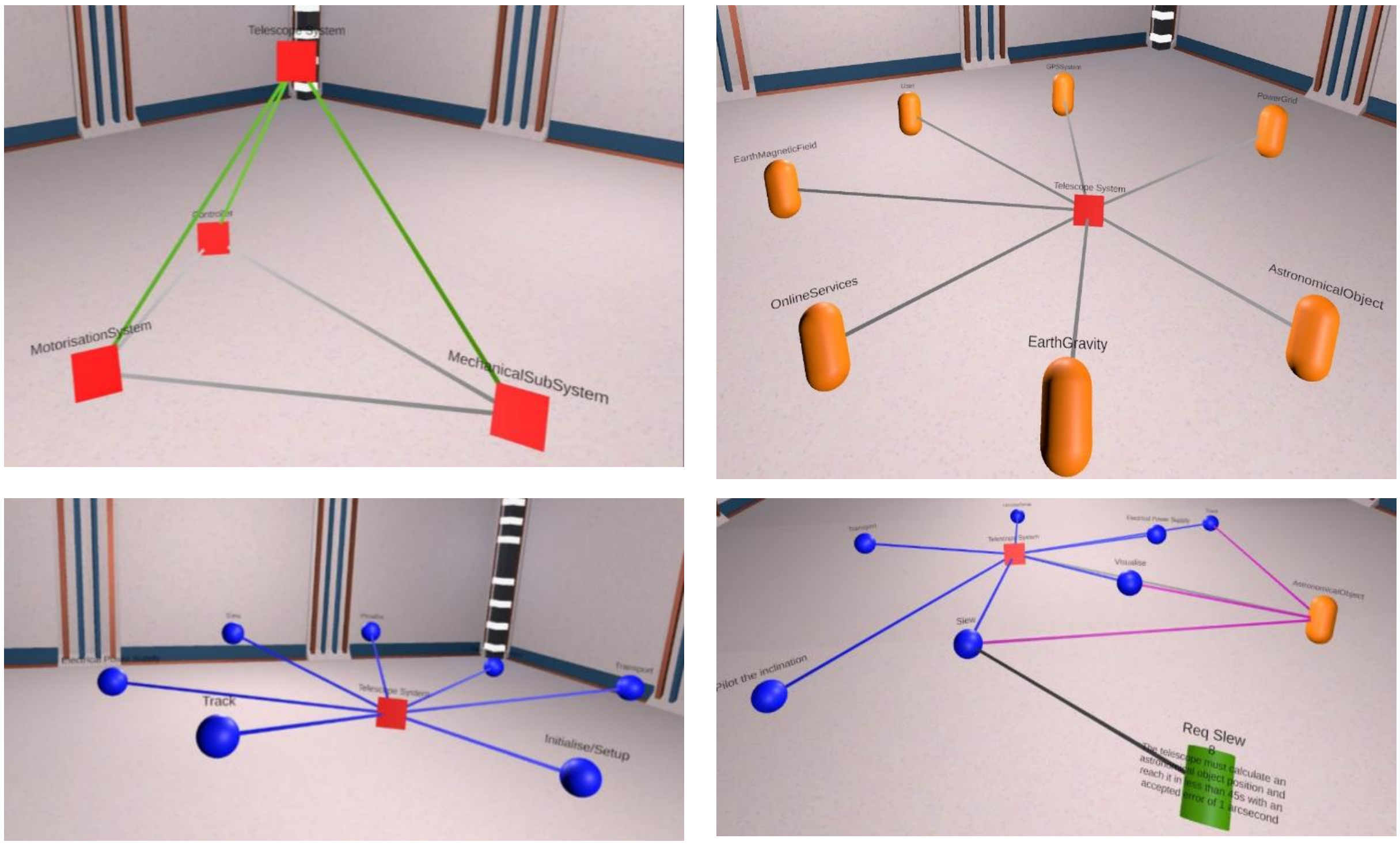

**Fig. 4.** Concepts encoded by the 3D Objects: red cube = building block (system or subsystem); blue sphere = function; orange ellipsoid = external entity; green cylinder = requirement

- **[Cognitive integration and Complexity management]** A visual notation should include visual cues that aid the cognitive integration process. This cognitive integration process involves a "*perceptual integration process characterised by a purposeful search behaviour of transitioning between diagrams and a conceptual integration process characterised by a hypothesis generation and refinement behaviour*" [72]. GraphXplore significantly enhances the perceptual and conceptual integration processes, as navigating and transitioning between diagrams is no longer necessary. All relevant visual items, typically dispersed across various diagrams from different perspectives, are naturally linked in the 3D graph summary (long shot) representation, which offers a view of the system as a whole. To reduce complexity management, GraphXplore displays each concept on a separate layer, avoiding overloading the human mind.

Regarding performance considerations, the current architecture of GraphXplore efficiently handles the Go-To telescope dataset without any system latency perceived by the end user, maintaining the optimal frame rates required for comfortable immersion. As for rendering scalability, we have not yet attempted to create and render large graphs, as evaluating enterprise-scale datasets was not the intent of this foundational study. However, we acknowledge that scaling to display tens of thousands of nodes simultaneously could induce system latency, which in VR directly correlates to cybersickness. Future iterations addressing massive datasets will necessitate performance optimisation techniques, such as dynamic occlusion culling, level-of-detail rendering, and graph-filtering algorithms, to maintain an acceptable frames-per-second threshold.

## 3.4. GraphXplore interactions

Virtual environment interactions fall into three tasks: viewpoint motion control, selection, and manipulation [73]. For viewpoint motion control or navigation, the virtual environment provides the most natural mode of travel: real walking, as user movement is confined to the room's area [74]. Users can employ the prevalent one-handed “Point and Teleport” navigation technique for longer distances, enabling them to teleport by simply pointing to a specific location before being instantaneously transported to those coordinates. This “Point and Teleport” navigation technique reduces the likelihood of collisions with obstacles compared to the Walk-in-Place technique, which utilises a joystick [75]. Additionally, it is generally preferred due to its lower demand on user effort.

The manipulation task enables the user to move the 3D objects with a virtual hand represented by a virtual controller (Fig. 5).

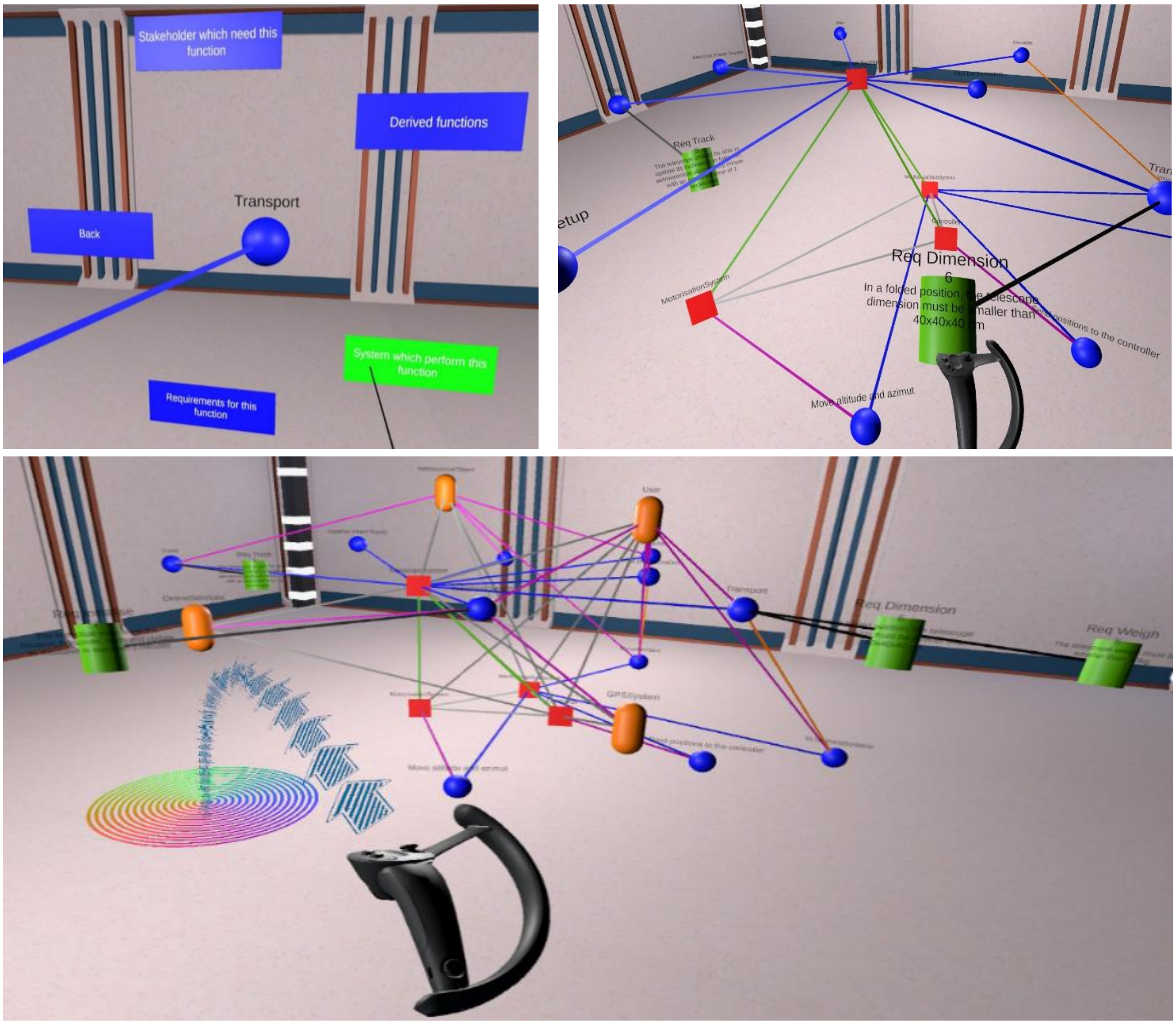

**Fig. 5.** Interactions for selecting (top left), manipulating (top right), and navigating (bottom) 3D objects in the virtual environment

A laser pointer interaction metaphor enables the user to select a 3D object. The user activates or deactivates the pointer and clicks on any node by pressing the trigger of the controller. Fig. 6 (left) illustrates the primary contextual menu (e.g., hide, develop, back) after the user pulls the trigger while aiming at a node. The hide option offers choices to remove a single node along with its adjacent edges or all nodes of a specific type from the scene, maintaining a comprehensible virtual environment by limiting the number of 3D objects. Developing a node (Fig. 6 right) enables the user to display its neighbourhood by selecting a data type (e.g., external entities who require functions, system(s) that perform a function, requirements derived from a function, etc.). When a submenu option is chosen, a predefined graph-oriented database query is sent to update the visualisation.

# 4. Empirical comparison of diagrams in slides with GraphXplore

As discussed in the literature review, empirical evidence does not support existing claims about the added value benefits of replacing desktop-based box-and-arrow diagrams with immersive human-model interactions or their integration. Our initial interest was to determine whether VR could help equip engineers with more natural and richer human-model interactions, thus improving collaboration between MBSE notational experts and non-experts.

The present study was designed to compare MBSE representations embedded in a PowerPoint slide deck against an immersive layered 3D graph for reviewing model-based system design. To achieve this comparison, we conducted a between-subjects controlled experiment in which participants responded to questions about exploring and understanding digital threads linking cross-domain data generated from a model-based systems engineering definition. To mitigate learning effects and confounding factors, such as fatigue and frustration, we employed a between-group design, exposing participants solely to either the immersive 3D graph (3D Graph) or the MBSE representations in a PowerPoint slide deck (Slide). This experimental design enables us to pose the same questions systematically regarding the comprehension of the MBSE definition. Furthermore, we assume that the interactive data exploration tasks required to answer the questions are straightforward and involve limited cognitive processes, thereby minimising individual differences, which makes a between-group design appropriate.

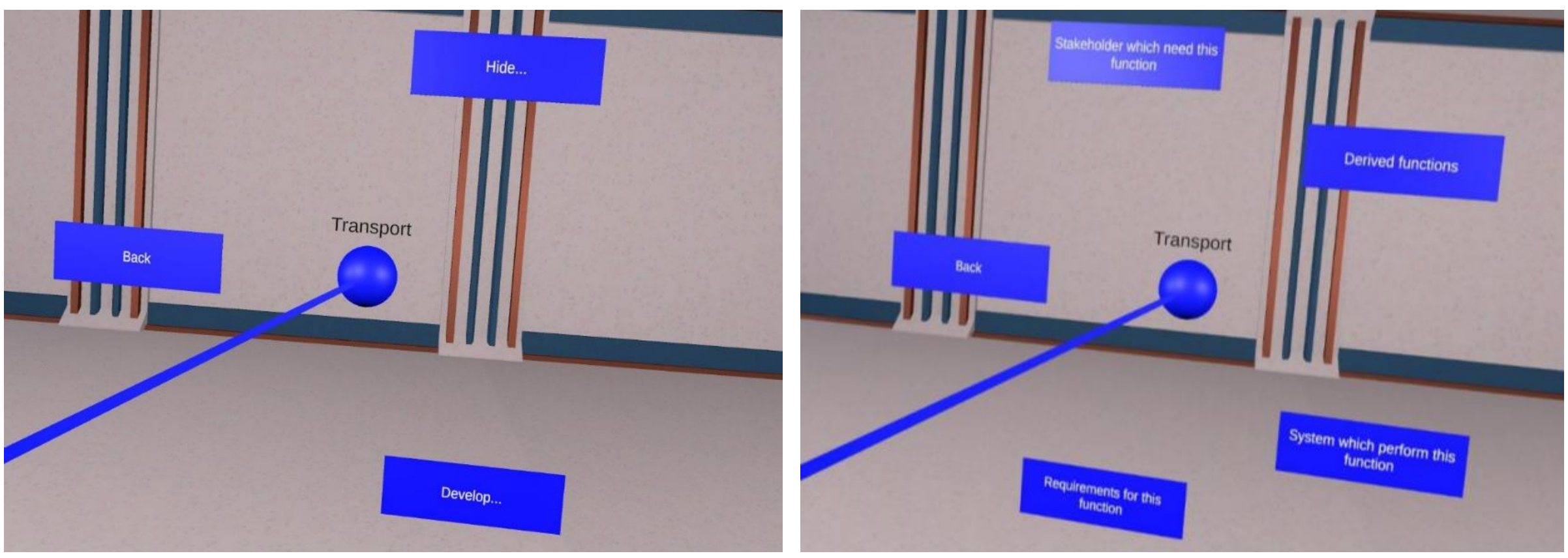

**Fig. 6.** Main contextual menu (left) and submenu (right) after selecting the option “Develop…”

## 4.1.Research questions and hypotheses

To compare the MBSE notations embedded in PowerPoint slides with the immersive layered 3D graph, we framed a pair of null hypotheses and alternative hypotheses for each research question derived from the more generic research questions stated in Section 1.4:

**Q1. Is the error score lower for the 3D Graph than for the Slides?**

- $H_0$: There is no difference in the error score between the 3D Graph and the Slides.
- $H_1$: The 3D Graph will be less error-prone than the Slides.

There is typically a trade-off between efficiency and accuracy, which means that when other factors remain constant, achieving a higher speed will lead to more errors, while ensuring fewer mistakes will reduce speed. Therefore, this study also measures how quickly participants complete the tasks to avoid overlooking a critical aspect of the picture.

**Q2. Is the completion time lower for the 3D Graph than for the Slides?**

- $H_0$: There is no difference in the completion time between the 3D Graph and the Slides.
- $H_1$: The completion time is lower with the 3D Graph than with the Slides.

**Q3. Is the perceived confidence higher for the 3D Graph than for the Slides?**

- $H_0$: There is no difference in the perceived confidence between the 3D Graph and the Slides.
- $H_1$: The perceived confidence is higher with the 3D Graph than with the Slides.

**Q4. Is the recall score higher for the 3D Graph than for the Slides?**

- $H_0$: There is no difference in the recall score between the 3D Graph and the Slides.
- $H_1$: Individuals who use 3D Graph have a better recall score than those who use Slides.

**Q5. Is the perceived usability higher for the 3D Graph than for the Slides?**

- $H_0$: There is no difference in perceived usability between the 3D Graph and the Slides.
- $H_1$: The perceived usability of the 3D Graph is higher than the Slides.

**Q6. Is the perceived workload lower for the 3D Graph than the Slides?**

- $H_0$: There is no difference in the perceived workload between the 3D Graph and the Slides.
- $H_1$: The perceived workload is lower for the 3D Graph than the Slides.

## 4.2. Ethics approval

The local ethics committee at the University of Grenoble Alpes approved the study. The trial was registered prior to the commencement of the study. All procedures were conducted in accordance with the ethical standards outlined in the 2013 Helsinki Declaration.

## 4.3. Experiment protocol

After designing the experiment (Sections 4.4 and 4.5), conducting pilot studies (Section 4.8), and recruiting participants (Section 4.6), the experimental protocol began with presenting the context and goals of the experiment to each volunteer participant. We briefly explained what a design review is, its different objectives, and the main decisions made during such reviews. Subsequently, we clearly informed them that our focus lies in understanding a system design definition, which is a fundamental objective for various design review goals, including completeness and correctness checking, change propagation analysis, and design trade studies. Before commencing the experiment, all participants provided their written informed consent, which had received ethical approval from the University Grenoble Alpes ethics committee prior to testing. This was followed by a pre-experimentation questionnaire that gathered demographic data, perceptions of the value of VR, and levels of experience in functional, behavioural, and structural product design. Each participant was assigned to a testing environment in a completely random manner using a simple random assignment, without stratifying for prior VR experience, ensuring that the researchers had no influence, either intentional or subconscious, on the treatment of the participant. A brief demonstration of the testing environment then took place, where the moderator illustrated how to set up the HMD and perform the main interactions (navigation, selection, manipulation) before allowing the participant to set up the HMD, execute the interactions, and ask any questions. For the controlled group, participants were required to review model screenshots stored in a PowerPoint slide deck. Following this brief demonstration, a training phase commenced to ensure that participants were familiarised with the modality and Model-Based Systems Engineering tasks. Using the training case study detailed in Section 4.8, the participant engaged in a training process involving a task akin to the main task and a dataset with the same characteristics but slightly reduced in size. The participant validated the training phase by successfully answering the training research questions after one or more attempts. This was followed by the experimental phase, during which the participant had to reuse the testing environment practised during the training phase to explore a new MBSE definition and respond to six questions related to a new Go-To telescope case study. Finally, the participant completed exit questionnaires on usability and cognitive load to conclude the experiment.

## 4.4. Independent variable

The unique controlled independent variable is the testing environment, which includes two visualisation conditions: the 3D Graph and the Slides. Both testing environments were assessed using the same case study, a Go-To telescope, selected to represent a real-world industrial electro-mechanical system. Each discipline, including functional analysis, offers its perspective on the system, materialised through various models developed with domain-specific modelling languages and software (Fig. 7). The first model aligns with the functional view and comprises 17 SysML diagrams that define use cases, contexts, functions, requirements, and architectures at both the system and subsystem levels. All SysML diagrams were created using the Papyrus modelling software. The second model, representing the behavioural view, delineates logical units (such as a command, a controller, a DC motor, and a multi-body structure) using Modelica and Wolfram System Modeler as the systems modelling language and software, respectively. The third model captures the structural view, which includes a preliminary CAD assembly of two solid parts: the mechanical support and the optical tube. The final model type features traceability matrices linking data from different views (e.g., architecture to CAD). A Go-To telescope was also preferred, as it was highly unlikely that participants were familiar with the system's design and, consequently, could answer the questions accurately.

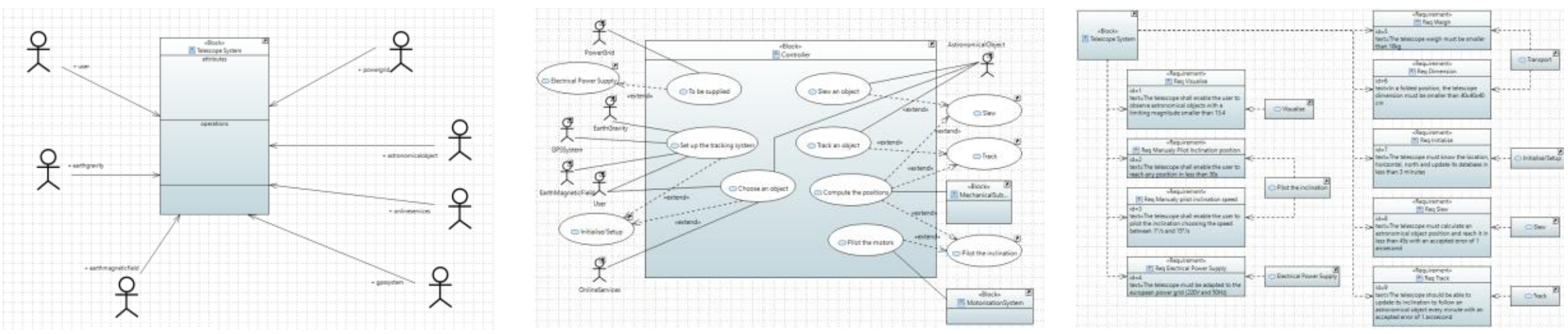

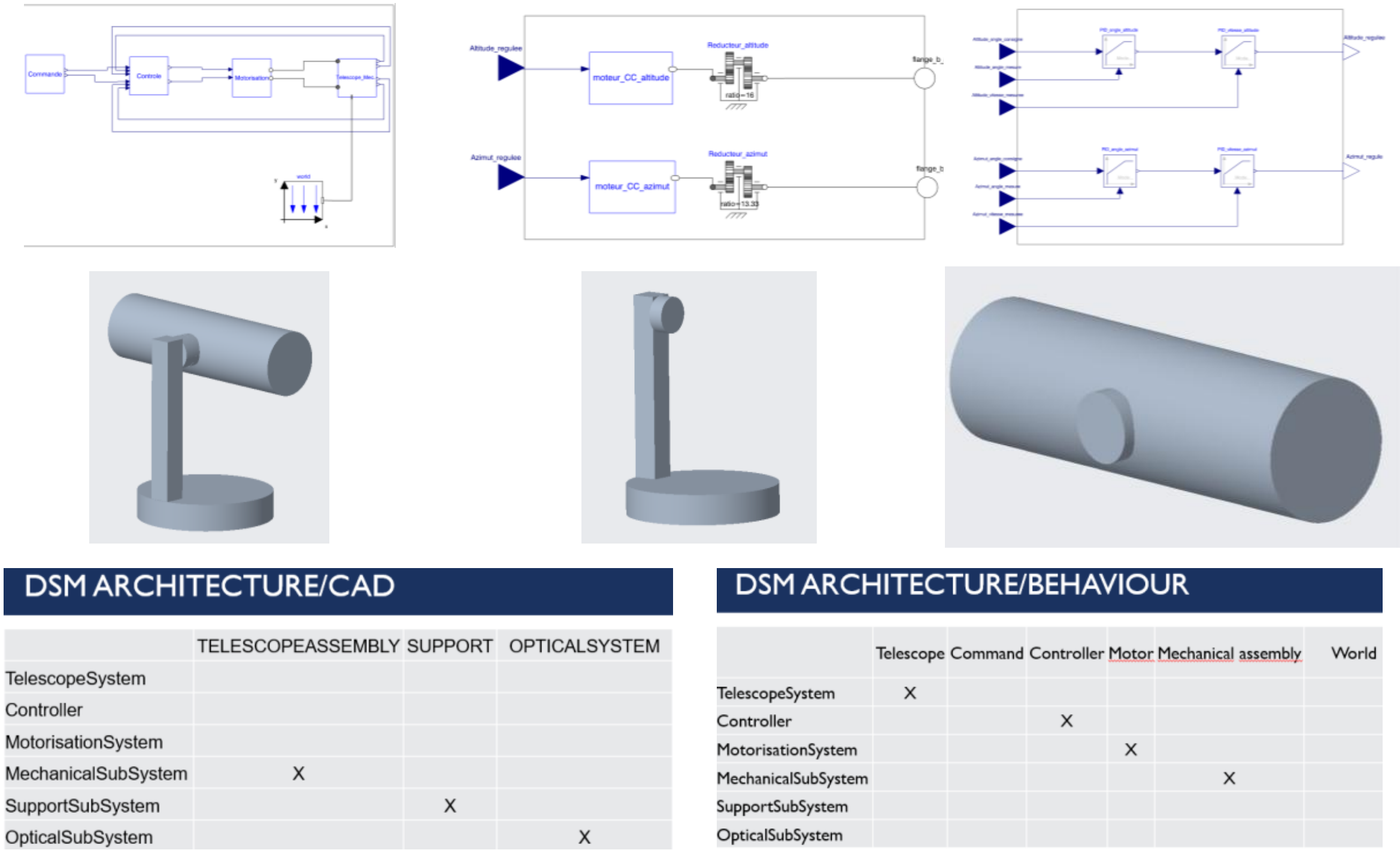

DSM ARCHITECTURE/CAD

| | TELESCOPEASSEMBLY | SUPPORT | OPTICALSYSTEM |
|---|---|---|---|
| TelescopeSystem | | | |
| Controller | | | |
| MotorisationSystem | | | |
| MechanicalSubSystem | X | | |
| SupportSubSystem | | X | |
| OpticalSubSystem | | | X |

DSM ARCHITECTURE/BEHAVIOUR

| | Telescope | Command | Controller | Motor | Mechanical assembly | World |
|---|---|---|---|---|---|---|
| TelescopeSystem | X | | | | | |
| Controller | | | X | | | |
| MotorisationSystem | | | | X | | |
| MechanicalSubSystem | | | | | X | |
| SupportSubSystem | | | | | | |
| OpticalSubSystem | | | | | | |

**Fig. 7.** Illustrations of models defining the functional (1st row), behavioural (2nd row), structural (3rd row) views, and traceability matrices (4th row) of the Go-To telescope case study used in the experiment

## 4.5. Dependent variables

Objective performance measures include the error score, recall score, and completion time. Subjective measures are perceived confidence, perceived usability, and perceived cognitive load.

To evaluate the error rate, participants were asked to answer three questions (Table 3) about the MBSE definition of the Go-To telescope. The error score refers to the raw number of mistakes made while answering Q1, Q2, and Q3, without considering the total number of trials, as we are interested in the absolute number of mistakes made by each participant. An error constitutes a false negative (when an item from the list of expected responses is missing) or a false positive (when an item is in excess). The lower the error score, the better the comprehension of the Go-To telescope MBSE definition.

**Table 3** Questions to measure the error score related to understanding the Go-To telescope MBSE definition

| | |
|---|---|
| ***Q1*** | What is the ID of the requirements satisfied by the subassembly “Support” from the structural view? |
| ***Q2*** | What is the name of the function(s) linked to the external entity “PowerGrid”? |
| ***Q3*** | What is the name of the external entity(ies) at the origin of the requirement “To manually pilot inclination position”? |

To measure perceived confidence, participants indicated their level of confidence in their answers on a 7-point Likert scale, ranging from 1 (not at all confident) to 7 (extremely confident), after answering each question.

The time performance refers to the duration required to answer questions 1, 2, and 3, beginning with the appearance of question 1 and concluding with the answer to question 3. This includes the time taken to search for information by exploring the models, all within a 20-minute limit. Questions were presented one after another to prevent participants from answering all the questions at once. Indeed, by scrolling sequentially through the questions, we avoid giving participants with strong memories an advantage, as they cannot memorise all the questions and answer all of them simultaneously. This approach helps maintain a conventional design review scenario, where engineers typically proceed sequentially to address a list of exit questions individually. We also examined the trade-off between error score and completion time. The NASA Task Load Index (TLX) [76] and System Usability Scale (SUS) [77] questionnaires measure perceived cognitive load and perceived usability, respectively.

After answering questions 1, 2, and 3, participants were invited to spend 5 minutes exploring the data within their testing environment to study the Go-To telescope MBSE definition. Finally, to assess the impact of the environment on the recall score, participants were asked to answer three additional questions (Table 4) without access to the models after completing the free exploration.

**Table 4** Questions to measure the recall score related to the understanding of the Go-To telescope MBSE definition

| | |
|---|---|
| ***Q4*** | What are the external entities of the telescope? |
| ***Q5*** | What are the logical blocks (of the behaviour view) composing the telescope? |
| ***Q6*** | What external entities are linked to the sub-system “MechanicalSubSystem” (in the architecture model)? |

## 4.6.Participants

Eligibility criteria for participants include a background in engineering design, self-reported normal vision, and fluency in either French or English.

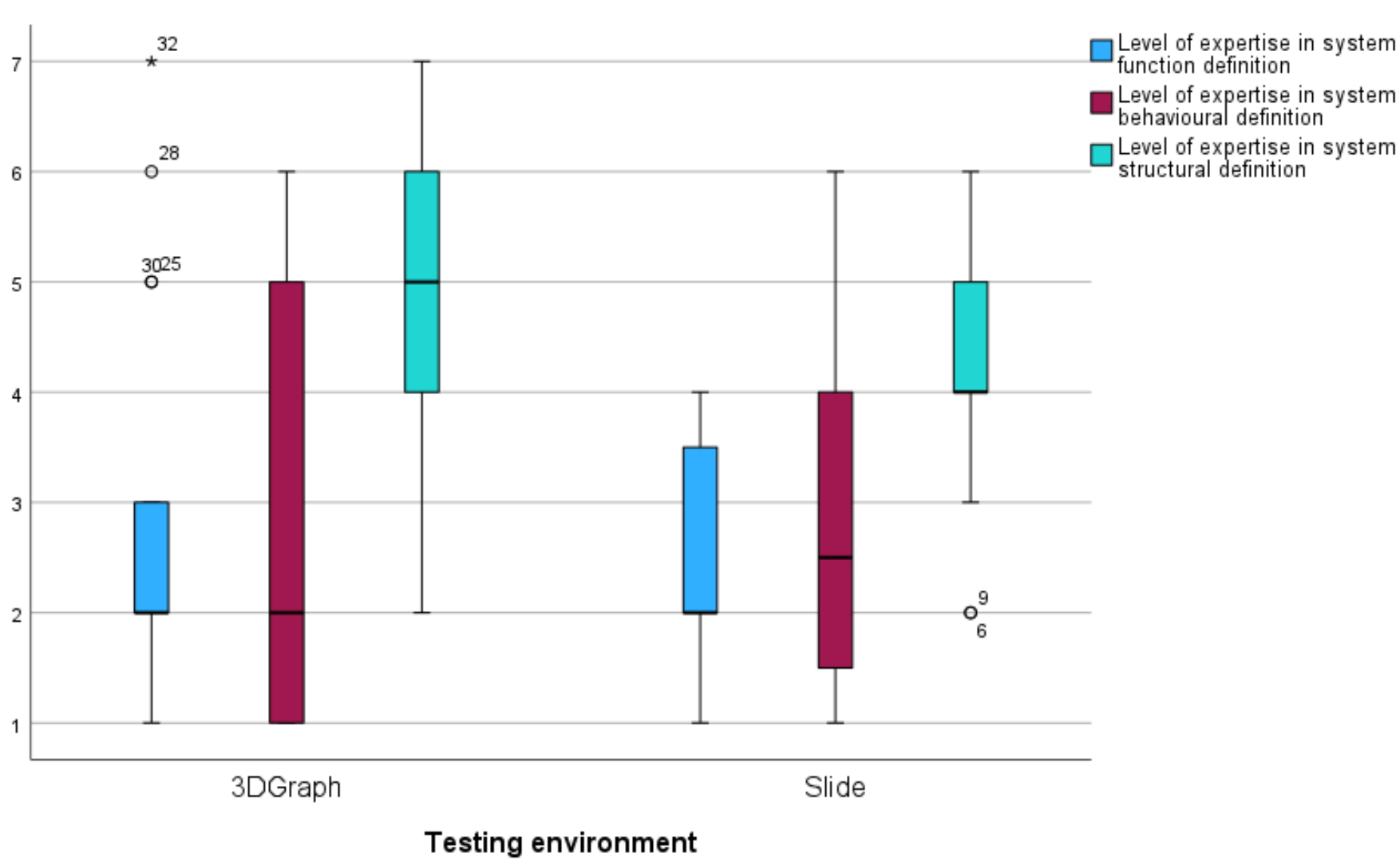

**Fig. 8.** Box plot summarising the level of experience of the participants in using models

A total of 33 participants were recruited from the students and researchers of the University of Grenoble Alpes. Their volunteer participation was motivated by an open call during the systems engineering lectures and an email sent to all Grenoble Institute of Technology students, professors, and the G-SCOP research lab members. The mean age of the participants was 26 (min: 20; max: 46; mean: 26.2; SD: 7.1). They included six women and 27 men. The control group consisted of 3 females (min: 20; max: 22; mean: 20.6; SD: 1.1) and 13 males (min: 20; max: 46; mean: 27; SD: 7.8), and the VR group consisted of 3 females (min: 21; max: 25; mean: 22.7; SD: 2.1) and 14 males (min: 20; max: 43; mean: 27.3; SD 7.5). Each participant had normal or corrected-to-normal vision.

Before starting the experimentation, participants were asked to qualitatively evaluate themselves on a scale of 1 (novice) to 7 (expert) regarding their experience with using engineering models and VR tools, as well as their opinions on the efficiency of VR for industrial practices.

Fig. 8 shows that the majority of participants ($N_{3D\ Graph}$ = 17, $M_{3D\ Graph}$ = 4.71, $SD_{3D\ Graph}$ = 1.532; $N_{Slide}$ = 16, $M_{Slide}$ = 4.31, $SD_{Slide}$ = 1.250) are comfortable with the use of design models defining the structural view of a system but have limited experience with models describing the functional ($M_{3D\ Graph}$ = 2.94, $SD_{3D\ Graph}$ = 1.784; $M_{Slide}$ = 2.44, $SD_{Slide}$ = 1.094) and behavioural ($M_{3D\ Graph}$ = 2.88, $SD_{3D\ Graph}$ = 1.833; $M_{Slide}$ = 2.75, $SD_{Slide}$ = 1.528).

Regarding VR, Fig. 9 shows that although participants have rarely experienced VR technologies (M = 2.71, SD = 1.61), a majority of those in the 3D Graph group believe it can be helpful (M = 5.29, SD = .985).

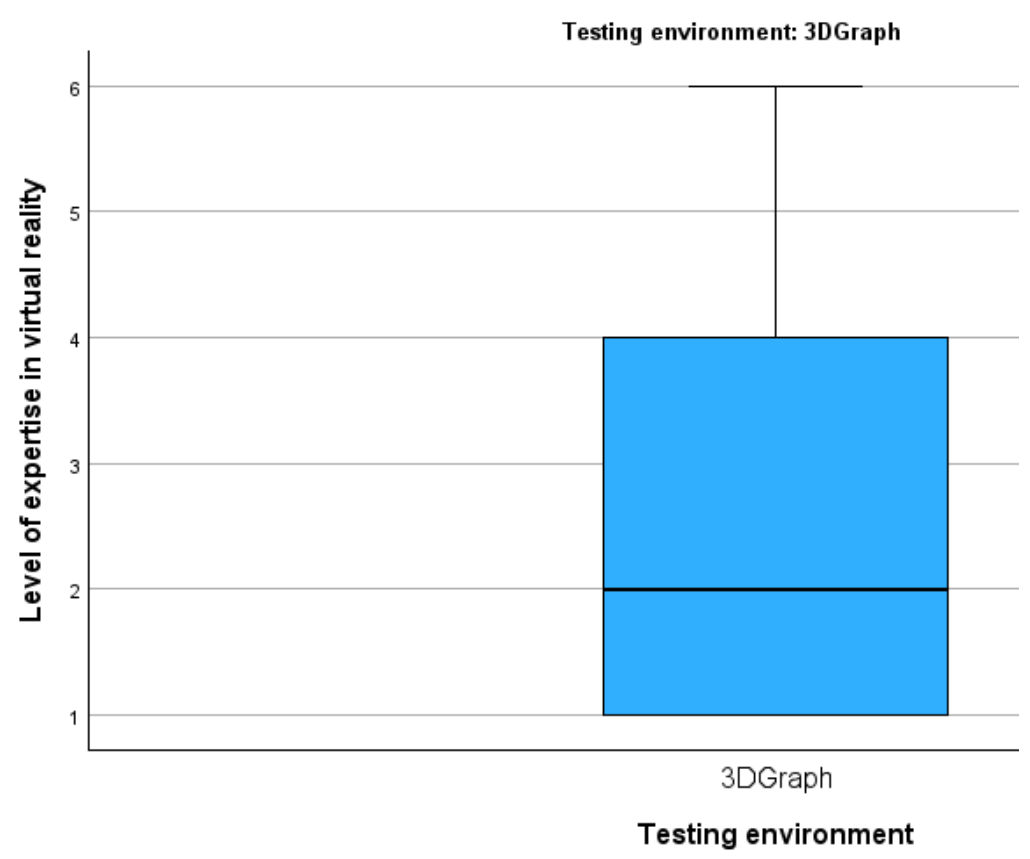

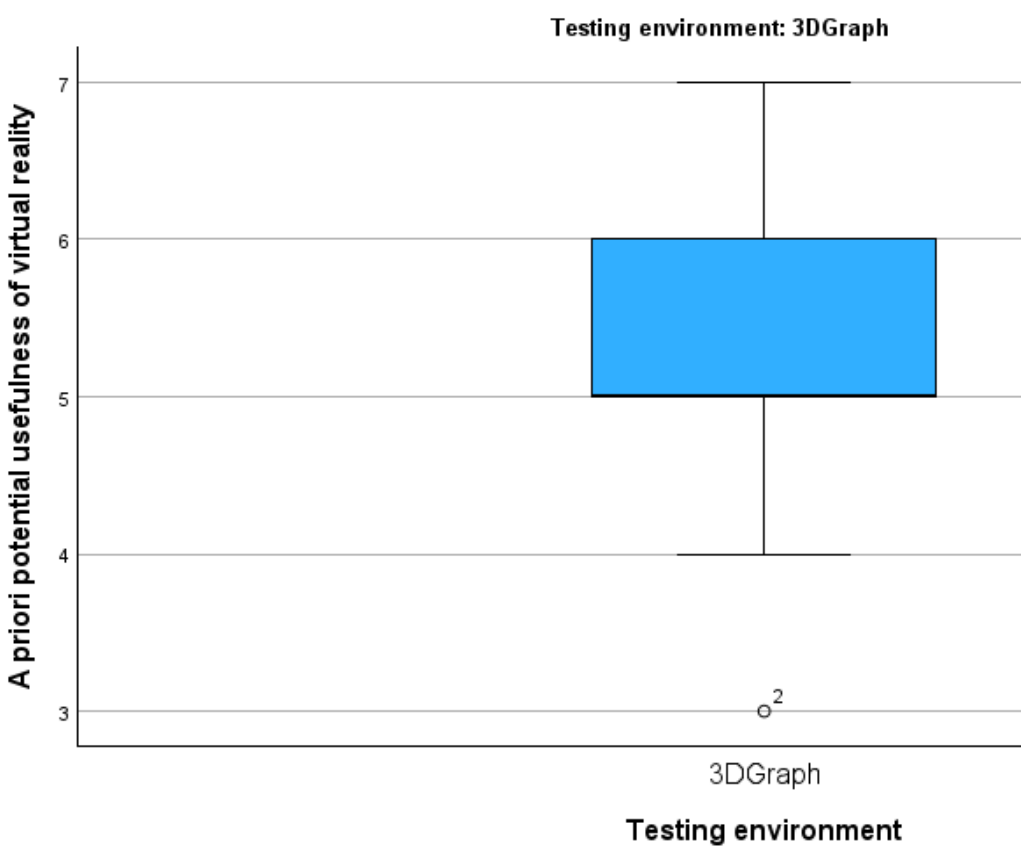

**Fig. 9.** Box plot summarising the level of experience in VR (left) and a priori usefulness of VR (right)

## 4.7. Apparatus and set up installation

Throughout the experiment, participants accessed two PCs (Fig. 10). The first PC was utilised to read and respond to the questionnaires. The second PC provided access to the assigned testing environment. All participants assigned to GraphXplore used a Valve Index headset.

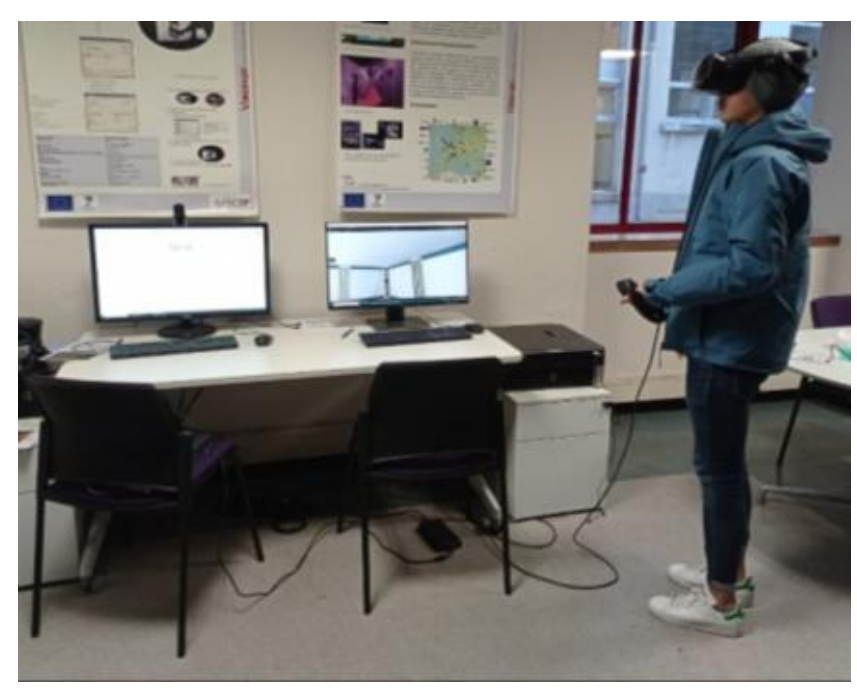

**Fig. 10.** Experimental setup with PC 1(left) to answer the questions and PC 2 (right) to use the testing environment

## 4.8. Pilot study

Before conducting the experiment, pilot studies were conducted to validate the experimental design. These studies aimed to ensure that the experiment would last less than an hour and that the questions posed to evaluate the error rate were understandable. Since the virtual environment necessitates training to learn how to use it, participants experimenting with it would require more time than those using the MBSE representations presented in a handcrafted PowerPoint slide deck. Therefore, to evaluate the duration of the experiment, the pilot studies were conducted using the VR environment. The recruited participants had no prior experience with VR environments. After a 20-minute training phase, the measured exercise commenced. Seven questions needed to be answered within the environment. Finding the answers compelled the user to explore the environment, which lasted 1 hour and 15 minutes. Following 5 minutes of free exploration, the participant answered three new questions without the aid of the environment to measure the recall score. This step took less than 10 minutes. Pilot studies lasted too long to be replicated on a larger scale, primarily because engaging participants for over an hour proved challenging. Consequently, four questions were removed from the pilot task, leaving only the most straightforward ones. Additional pilot studies were conducted with the redesigned experiment lasting less than an hour and validating the protocol. Results from the final design of the pilot studies have been included in the results, as the protocol remains the same for subsequent participants.

## 4.9. Training

To train participants on the task and the environment, the case study differs from the GO-TO telescope in that it avoids learning effects. We designed another case study corresponding to the model-based design of an inverted pendulum (Fig. 11). An inverted pendulum was chosen because, like the GO-TO telescope, it is an electromechanical system, but it is relatively simpler to analyse as it contains fewer model elements. The datasets used in the training and experimental exercises used the same data model (Fig. 3).

The control group was trained to familiarise participants with the task, without PowerPoint instruction, since they already knew it. VR group participants learned to navigate and interact in the virtual environment, with the investigator using a script to ensure consistent information and encourage questions.

Participants had 5 minutes to freely explore the model-based design of the inverted pendulum system before answering 3 questions (Table 5). This set of questions is similar to the questions asked in the experiment. The training of participants was validated when they correctly answered the three questions. During this training phase, participants were encouraged to ask questions about the task, the data and the environment used. In contrast, during the experimental phase, participants received no support from the researchers.

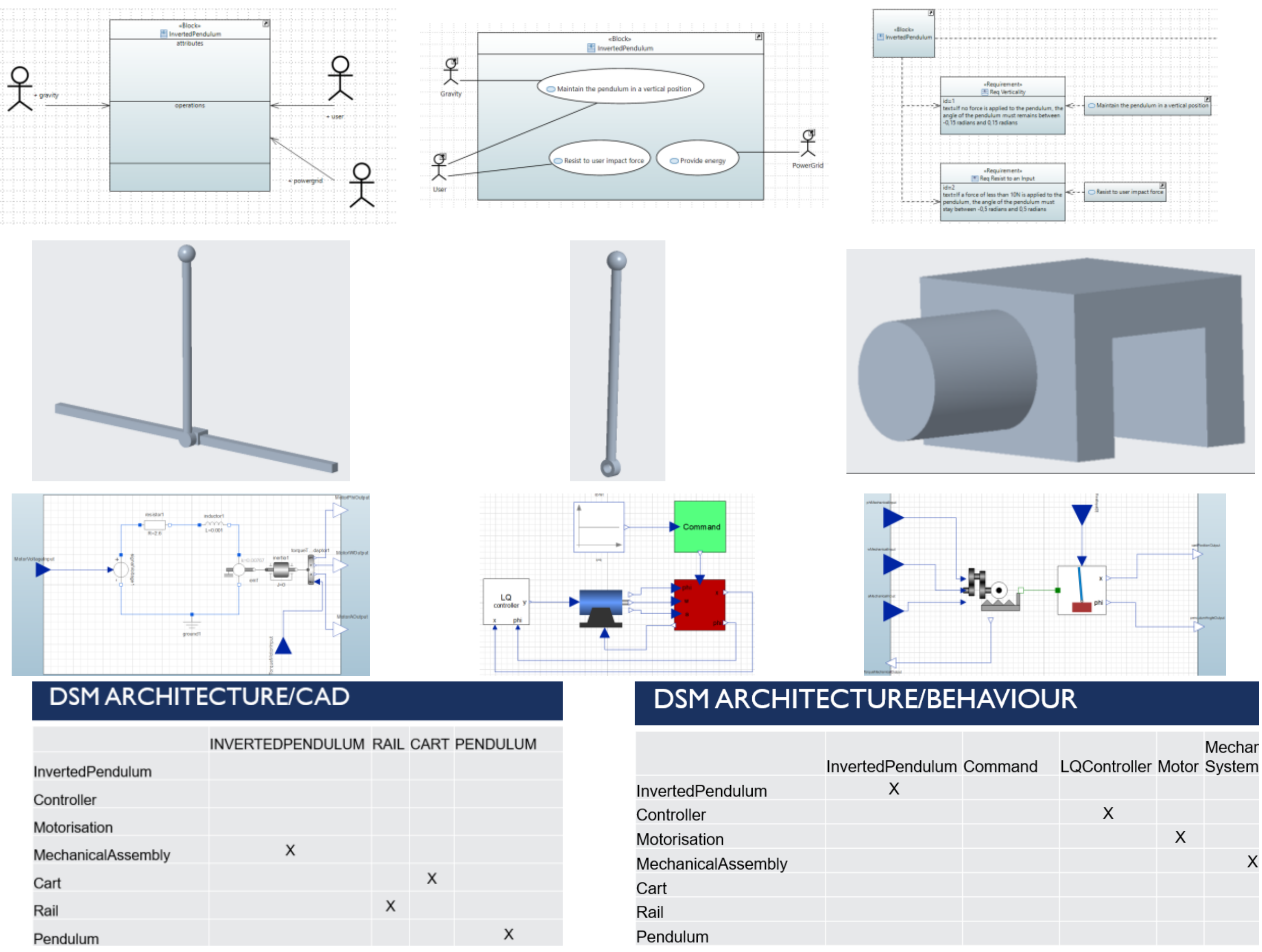

**Fig. 11.** Illustrations of models defining the functional (1st row), behavioural (2nd row), structural (3rd row) views, and traceability matrices (4th row) of the inverted pendulum case study used in the training

**Table 5** Questions to measure the error score related to the understanding of the inverted pendulum MBSE definition during the training phase

| | |
|---|---|
| ***Q1*** | What is the name of the requirement(s) for which the external entity “Earth gravity” is at the origin? |
| ***Q2*** | What is the name of the logical unit that satisfies the requirement named “ReqPilotMotor”? |
| ***Q3*** | What is the name of the function(s) performed by the assembly “INVERTEDPENDULUM_ASM”? |

## 4.10. Post-experiment questionnaires

For the last phase of the experiment, the participants must answer two questionnaires. The first one, the SUS, measures the usability of the environment. The second questionnaire, the NASA/TLX, measures the cognitive load of the task performed by the participants. Here, the task is answering the first set of three questions within a limited time of 20 minutes. These questionnaires evaluate the usability of GraphXplore and compare it with the conventional hand-crafted PowerPoint slide deck.

# 5. Results

## 5.1. Assumptions checking

Data from the participant responses were compiled into a single dataset, and the analysis of results started with checking the assumptions for an independent t-test. All dependent variables are measured on a continuous scale and consist of two categorical, independent groups. Observations are independent as each participant belongs to only one group, and there is no relationship between the observations in each group.

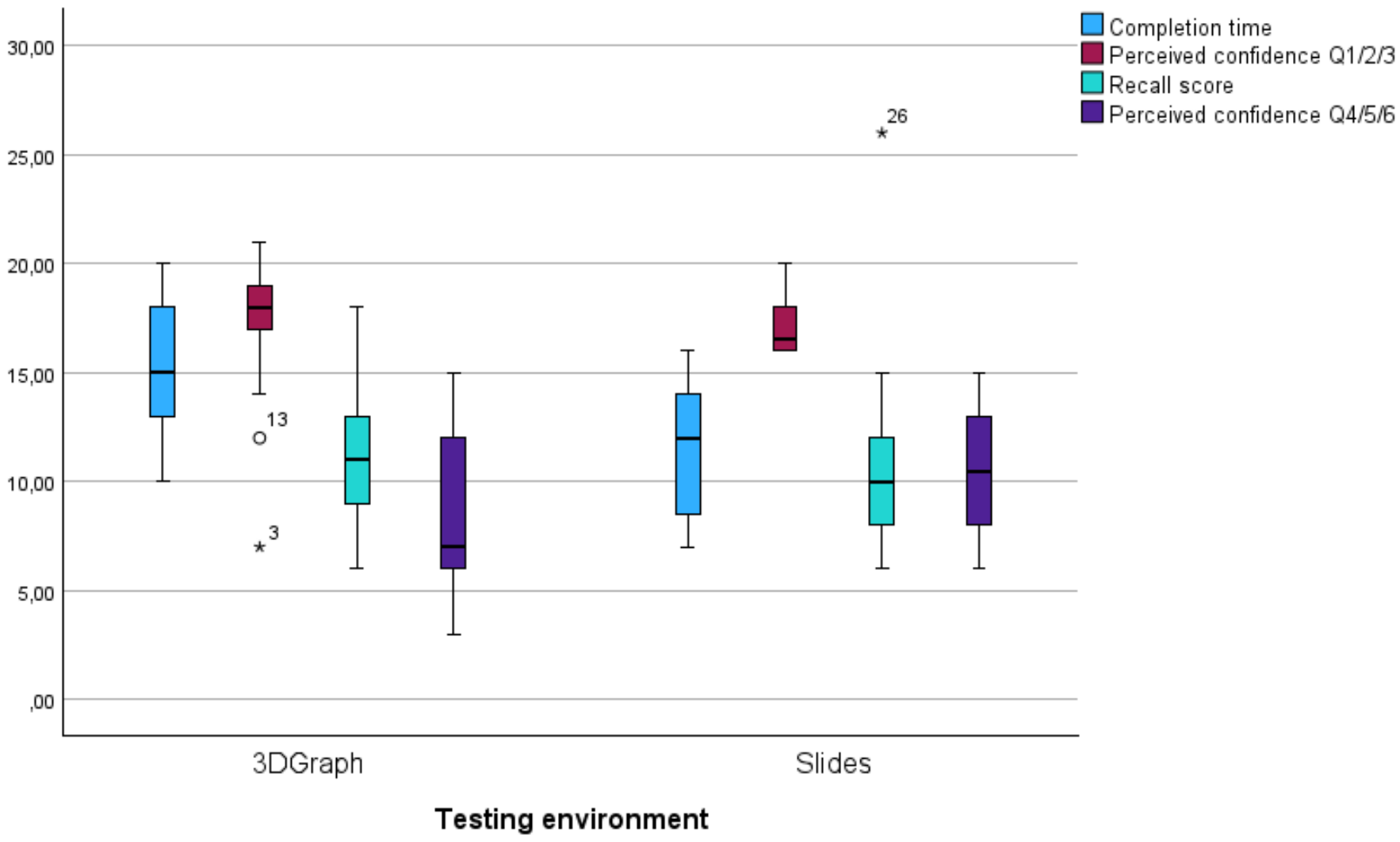

**Fig. 12.** Box plot of the completion time, perceived confidence on question Q1/2/3 and Q4/5/6, and recal score according to the testing environment

The following steps focused on identifying outliers, checking that the dependent variables are approximately normally distributed for each group of the independent variable, and the homogeneity of variances. Interquartile Range (IQR) was used to detect the candidate outliers (Fig. 12 without NASA-TLX and SUS scores). Once outliers had been detected, we tried to discern the cause (errors in the data, mischievous participants, faulty study design, or natural deviation). After a thorough check of the calculations, there are no errors in the data. In the 3D Graph group, outliers 12 and 13 answered questions 2 and 3 correctly but got question 1 completely wrong, which means they engaged appropriately (for your purposes) in the study, so we decided not to remove them. Participant 21 in the Slide group failed for a similar reason, as he scored poorly with 7 false negatives, 1 false positive for Q1 and 1 false negative for Q2. Therefore, we kept the results in our dataset. Regarding the extremely low recall score of Participant 26 in the control group, the items provided as answers to the open question are actual elements of the datasets and not random data. Confidence scores for Q4 (4 over 7), Q5 (3 over 7) and Q6 (1 over 7) are different too. Those elements suggest that the participant did what was expected, so we did not discard them.

Consequently, because our data includes non-parametric continuous variables and ordinal data (such as the Likert scales used for perceived confidence, SUS, and NASA-TLX), the Mann-Whitney U test was selected as the most robust statistical method for comparing the medians of the two independent groups. For continuous data that met the assumption of normality (e.g., completion time), we utilised an independent samples t-test.

**Table 6** reports the results of the analytical Shapiro-Wilk test to verify that the independent variables are approximately normally distributed for each group of the independent variable. We preferred the Shapiro-Wilk test to the Kolmogorov-Smirnov test because it is more appropriate for small sample sizes (< 50 samples). The 3D Graph and Slide samples were not normally distributed for all measures. When data does not conform to a normal distribution, statistical guidelines recommend applying a log10 transformation [78], but in our case, it did not help to make the data more normally distributed. Consequently, because our data includes non-parametric continuous variables and ordinal data (such as the Likert scales used for perceived confidence, SUS, and NASA-TLX), the Mann-Whitney U test was selected as the most robust statistical method for comparing the medians of the two independent groups. For continuous data that met the assumption of normality (e.g., completion time), we utilised an independent samples t-test.

**Table 6** Median, Mean (M), standard deviation (SD), and Shapiro-Wilk normality scores (Sig) for the dependent variables

| Criteria | 3D Graph | | | | Slides | | | |
|---|---|---|---|---|---|---|---|---|
| | *Median* | *M* | *SD* | *Sig* | *Median* | *M* | *SD* | *Sig* |
| Error score | 0 | 1.765 | 4.603 | <u><.001</u> | 0 | 1.062 | 2.515 | <u><.001</u> |
| Completion time | 15 | 15.471 | 3.044 | .277 | 12 | 11.375 | 3.030 | .265 |
| Efficiency | 0.080 | 0.173 | 0.334 | <u><.001</u> | 0.105 | 0.199 | 0.321 | <u><.001</u> |
| Perceived confidence Q1/2/3 | 18 | 17.178 | 3.609 | <u>.015</u> | 16.500 | 17.062 | 1.3889 | <u><.001</u> |
| Recall score | 11 | 11.177 | 3.321 | .496 | 10 | 10.938 | 4.697 | <u><.001</u> |
| Perceived confidence Q4/5/6 | 7 | 8.588 | 3.954 | .094 | 10.5 | 10.562 | 3.097 | .244 |
| SUS | 77.5 | 73.088 | 17.930 | .070 | 67.5 | 66.406 | 21.601 | .250 |
| NASA-TLX | 47.7 | 43.817 | 16.639 | .225 | 44.5 | 50.350 | 19.384 | .273 |

## 5.2. Hypotheses testing

The error score (Fig. 13) is not normally distributed, so a non-parametric Mann-Whitney U test was performed to evaluate whether the error score differed by the testing environment. The results indicated that the group with the 3D Graph (Mdn = 0) did not retrieve information significantly better than the group with the Slides (Mdn = 0), U = 132, p = .901, with a small effect size, r = .031. The mean rank for the 3D Graph group was 16.76, while the mean rank for the Slides group was 17.25.

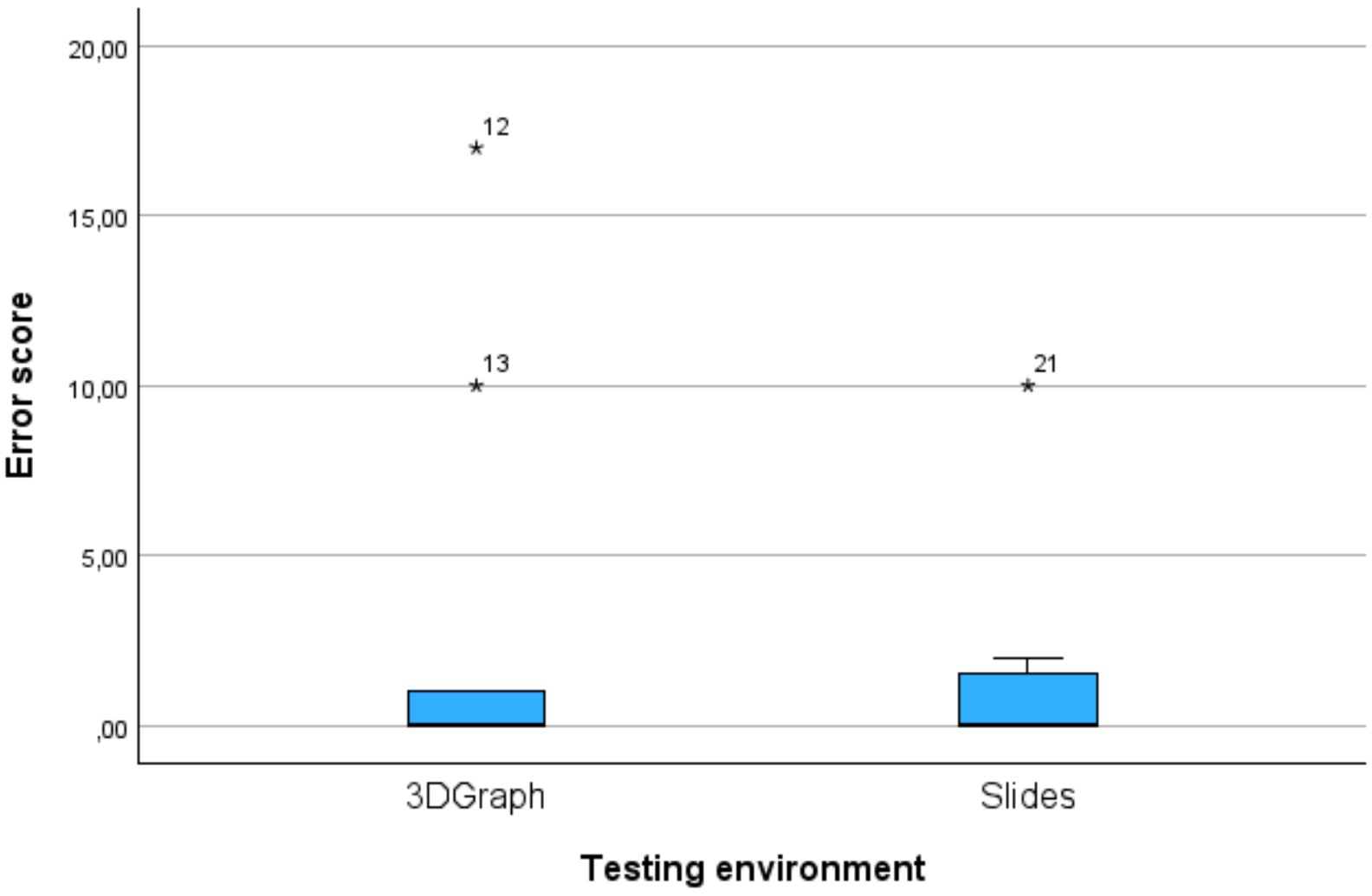

**Fig. 13.** Box plot of the error score according to the testing environment

The completion time is normally distributed, so an independent samples *t*-test was performed to evaluate whether there was a difference between the 3D Graph group and the Slides group. The results indicated that the 3D Graph group (*M* = 15.471, *SD* = 3.044) had a significantly higher completion time than the Slides group (*M* = 11.375, *SD* = 3.030), t(31) = 3.871, *p* = < .001. The effect size, as measured by Cohen's d, was d = 1.348, indicating a large effect. We used Cohen's d as both groups have similar standard deviations and sizes. This result indicates a significant difference in favour of the condition that used the Slides, which required less time.

The efficiency (Fig. 14) is not normally distributed, so a non-parametric Mann-Whitney U test was performed to evaluate whether the efficiency differed by the testing environment. The results indicated that the group with the 3D Graph (Mdn = 0.080) was not significantly more efficient than the group with the Slides (Mdn = 0.199), U = 87, p = .081, with a small effect size, r = .310. The mean rank for the 3D Graph group was 14.12, while the mean rank for the Slides group was 20.06.

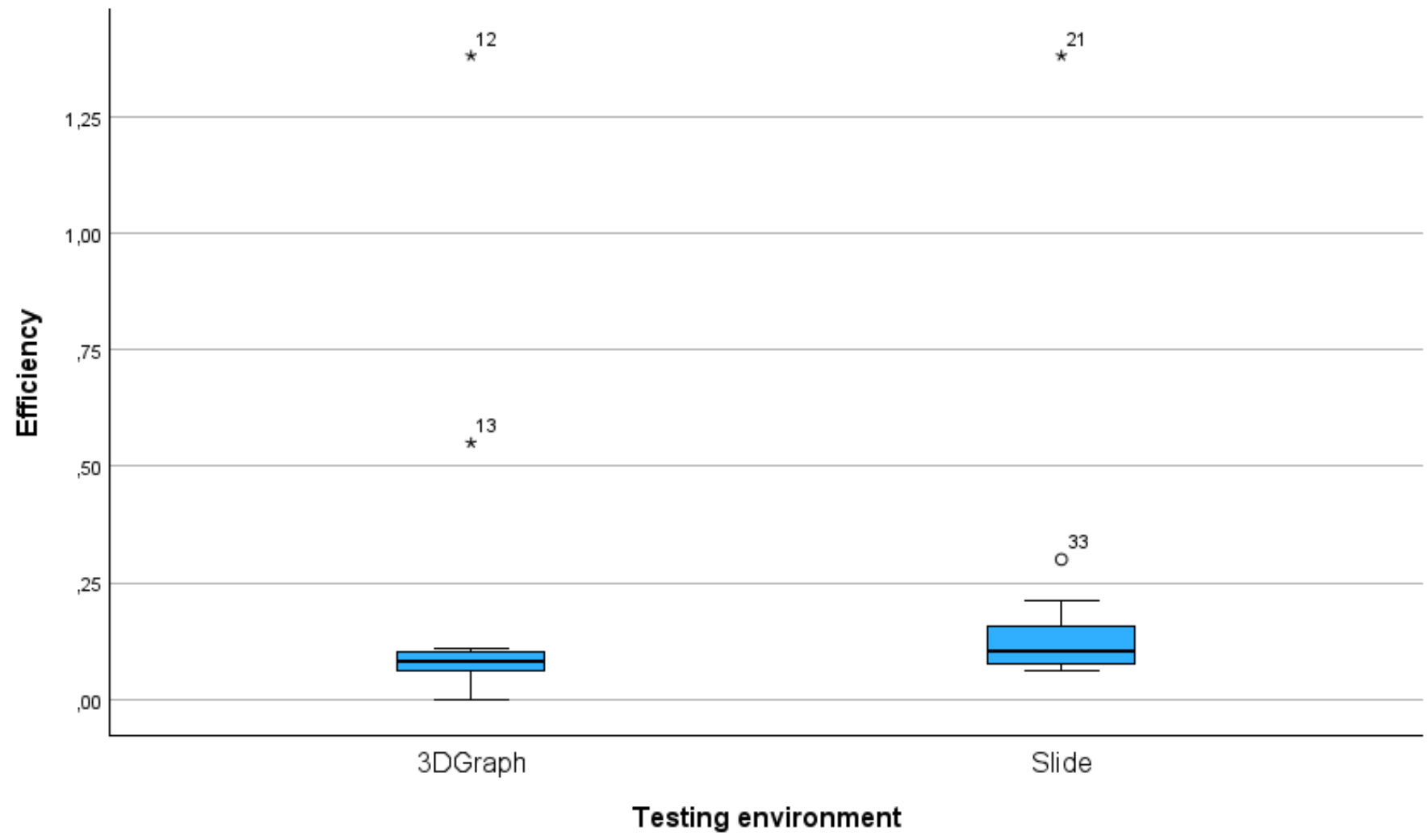

**Fig. 14.** Box plot of the efficiency according to the testing environment

The recall score (Fig. 15) is not normally distributed for the group with the Slides, so a non-parametric Mann-Whitney U test was performed. The results indicated that the group with the Slides (Mdn = 10) did not remember information significantly better than the group with the 3D Graph (Mdn = 11), U = 115, p = .465, with a small effect size, r = .132. The mean rank for the Slides group was 15.69, while the mean rank for the 3D Graph group was 18.24.

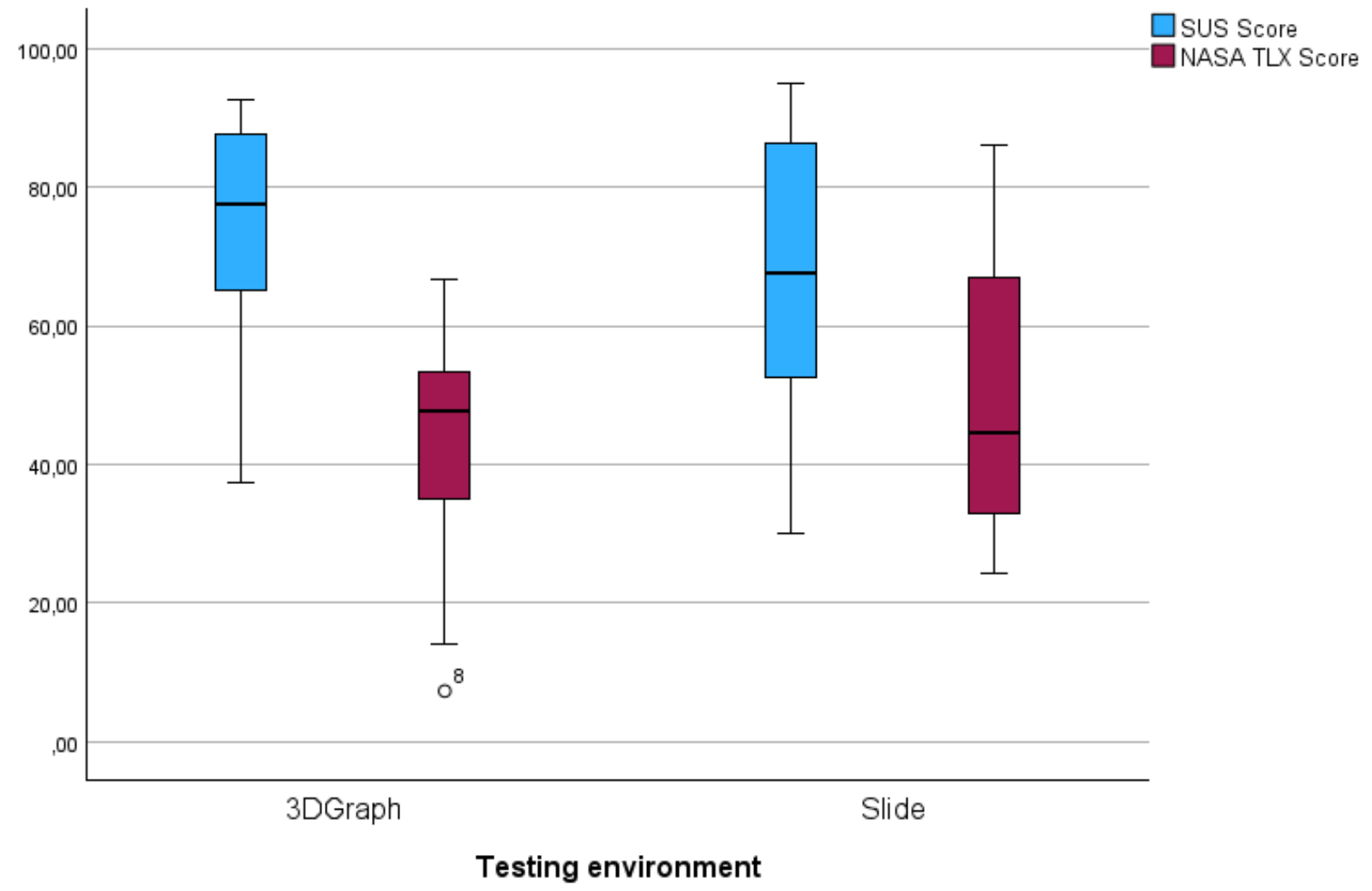

**Fig. 15.** Box plot of the SUS and NASA-TLX scores according to the testing environment

A non-parametric Mann-Whitney U test was performed to evaluate whether there was a significant difference in perceived confidence scores for Q1/2/3 and Q4/5/6 between participants using the 3D Graph and those using the Slides. The results indicated that the group with the 3D Graph (Mdn = 18) was not significantly more confident for Q1/2/3 than the group with the Slides (Mdn = 16.5), U = 104, p = .260, with a small effect size, r = .204. The mean rank for the 3D Graph group was 18.88, while the mean rank for the Slides group was 15.00. As for the Q4/5/6 recall task, the group with the 3D Graph (Mdn = 7) was not significantly less confident than the group with the Slides (Mdn = 10), U = 91.5, *p* = .110, with a small effect size, r = .281. The mean rank for the 3D Graph group was 14.37, while the mean rank for the Slides group was 19.78.

Given that Likert scale data are ordinal in nature, the Mann-Whitney U test was again chosen to determine whether there is a difference in SUS and NASA-TLX scores between participants with slides and participants with the 3D Graph. The results revealed no significant difference between the SUS score of the 3D Graph group (Mdn = 77.5) and Slides group (Mdn = 67.5), U = 111.5, p = .382, with a small effect size, r = 0.015. The mean rank for the 3D Graph group was 18.44, while the mean rank for the Slides group was 15.47. As for the NASA-TLX score, there was also no significant difference between the 3D Graph group (Mdn = 47.7) and the Slides group (Mdn = 44.5), U = 117, p = .510, with a small effect size, r = 0.120. The mean rank for the 3D Graph group was 15.88, while the mean rank for the Slides group was 18.19.

# 6. Discussion and threats to validity

The results do not provide evidence that the 3D Graph was more efficient than the Slides in comprehending a model-based design, at least the proposed model-based design of the telescope used as a case study. Therefore, they go against our intuitions as we were convinced that the holistic view and interactive navigation capabilities proposed by the immersive layered 3D graph would prevent practitioners from transitioning between multiple models that are required for understanding the model-based design, to search for relevant data dispersed among different windows and integrate it into a coherent mental representation. Indeed, complex work domains and inherently limited display capabilities have a fairly direct impact on the ability of practitioners to conduct work effectively, as discussed by [79]:

> "*All of the information required for effective control cannot possibly be displayed in parallel; the practitioner must selectively view glimpses serially, over time, through the limited ''keyhole'' provided by the small display surface. Thus, information must often be remembered and mentally integrated across successive glimpses, an activity that practitioners are not particularly well equipped to handle. A direct consequence is that practitioners must navigate these complicated workspaces and databases to find information that is relevant for the task at hand.*"

However, most of our results corroborate similar studies [15–18] in software engineering, which showed that designers working in VR are not yet more efficient than working on traditional screen setups for designing software or understanding complex software designs.

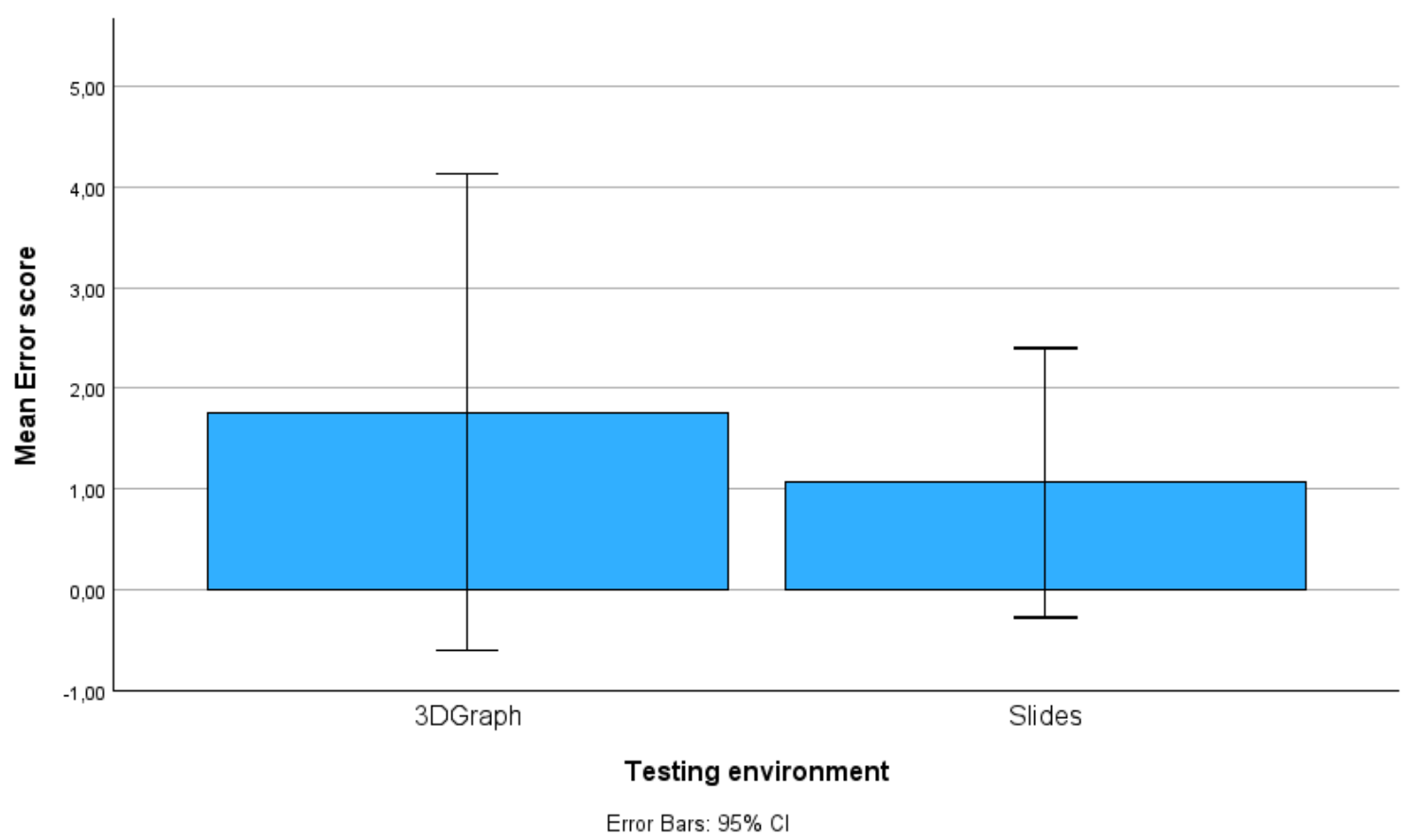

**Fig. 16.** Mean *error score* by testing environment (the lower the better). Error bars show 95% CI

Fig. 16 shows that the mean error score was slightly higher for the participants with the 3D Graph, but the consistently low error scores across both groups suggest that the tasks were too simple to fully measure the value of a holistic 3D graph summary. Furthermore, the Go-To telescope model, while representative of an electromechanical system, constitutes a relatively small and well-structured dataset. On a small graph of this scale, simply clicking through a few well-organised 2D slides remains a highly efficient method for basic information retrieval. The holistic, immersive visualisation offered by GraphXplore is arguably unnecessary for simple 'A-to-B' lookups, which explains why the 3D Graph did not significantly improve accuracy or completion time.

However, these results highlight a finding: despite the inherent overhead of learning to navigate a 6-DOF immersive environment, participants using GraphXplore reported higher perceived usability and lower perceived cognitive workload compared to the traditional slide-based setup. Proving that GraphXplore works smoothly and reduces cognitive friction on a foundational, small-scale dataset is a necessary prerequisite before scaling up. The true benefits of a holistic, immersive visualisation are hypothesised to emerge when dealing with massive, complex, and densely connected industrial datasets where 'getting lost' across fragmented 2D views becomes a severe cognitive bottleneck. Thus, establishing baseline usability and reduced mental demand here paves the way for future evaluations involving enterprise-scale models and complex conceptual integration tasks. However, as this study relied primarily on postgraduate students and researchers, generalising these results directly to industrial settings requires caution. While the fundamental information retrieval tasks used in this study did not strictly require deep domain-specific knowledge, the cognitive profile of the participant sample likely differs from that of senior professionals. Industrial practitioners possess deeply ingrained mental models and habits shaped by years of relying on traditional 2D MBSE interfaces (e.g., standard SysML diagrams, traceability matrices). Consequently, they might face higher initial adoption barriers and cultural resistance to the tool due to their established workflows. Furthermore, students and academic researchers often possess a higher baseline familiarity with 3D spatial navigation and immersive technologies, potentially reducing the initial cognitive load of using 6-DOF controllers compared to older practitioners. We acknowledge that recruiting a sufficiently knowledgeable sample of industrial practitioners to evaluate these complex sociotechnical and cultural adoption factors remains a significant practical challenge for future work.

The findings of improved usability and reduced cognitive workload carry implications for the future of collaborative design review practices. First, the reduction in perceived mental demand suggests that immersive 3D graphs could democratise the review process. By replacing rigid diagrams with an intuitive spatial topology, non-systems engineers, such as project managers, domain experts, or customers, can participate in collaborative sense-making without needing to master esoteric SysML syntax. Second, for immersive environments like GraphXplore to be effectively deployed in collaborative sessions, enterprises must fully commit to a federated, data-driven design approach. This necessitates establishing a global data schema, implementing robust data exchange mechanisms, and working closely with end-users to pre-define the specific design questions (e.g., change impact analysis, coverage analysis) and their corresponding database queries. Finally, while this study demonstrated the benefits of an HMD over a standard desktop setup, the physical environment of collaborative reviews must be considered. Future research must evaluate the specific influence of HMD immersion by comparing it against the large, high-resolution powerwall displays traditionally used in concurrent engineering rooms. Such a comparison is necessary to determine whether the reduced cognitive workload stems primarily from the stereoscopic 3D visualisation of large datasets or simply from the increased screen real estate.

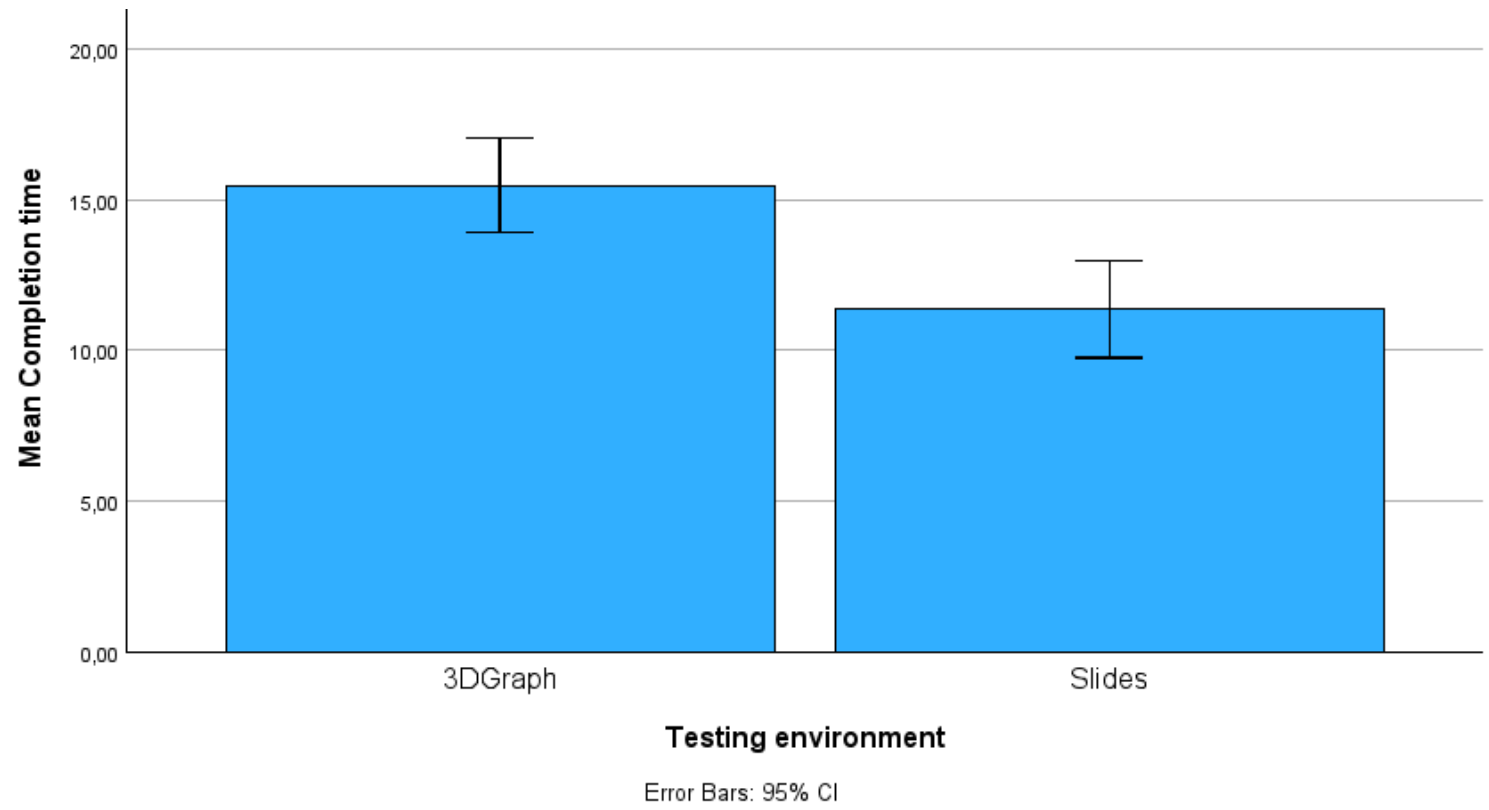

**Fig. 17.** Mean *completion time* by testing environment (the lower the better). Error bars show 95% CI

Fig. 17 shows that the group with the 3D Graph spent more time answering questions 1, 2, and 3 while interacting with the environment. The very low experience of participants with VR technologies (Fig. 9) and observations of their behaviours during the experiment motivate us to think that it is due to over-engagement with non-task-related elements (e.g., exploring the virtual room). Indeed, users are sometimes more interested in exploring the potential of the system. Physical discomfort with the VR HMD, which requires participants to readjust it, also negatively affects completion time. Conversely, as a user becomes more comfortable, they may become more engaged with the virtual environment and experience a stronger sense of immersion. This could increase their willingness to spend time in the environment, thués affecting the overall experiment duration. However, the training phase limited the learning curve and required time to adapt to the interactive system.

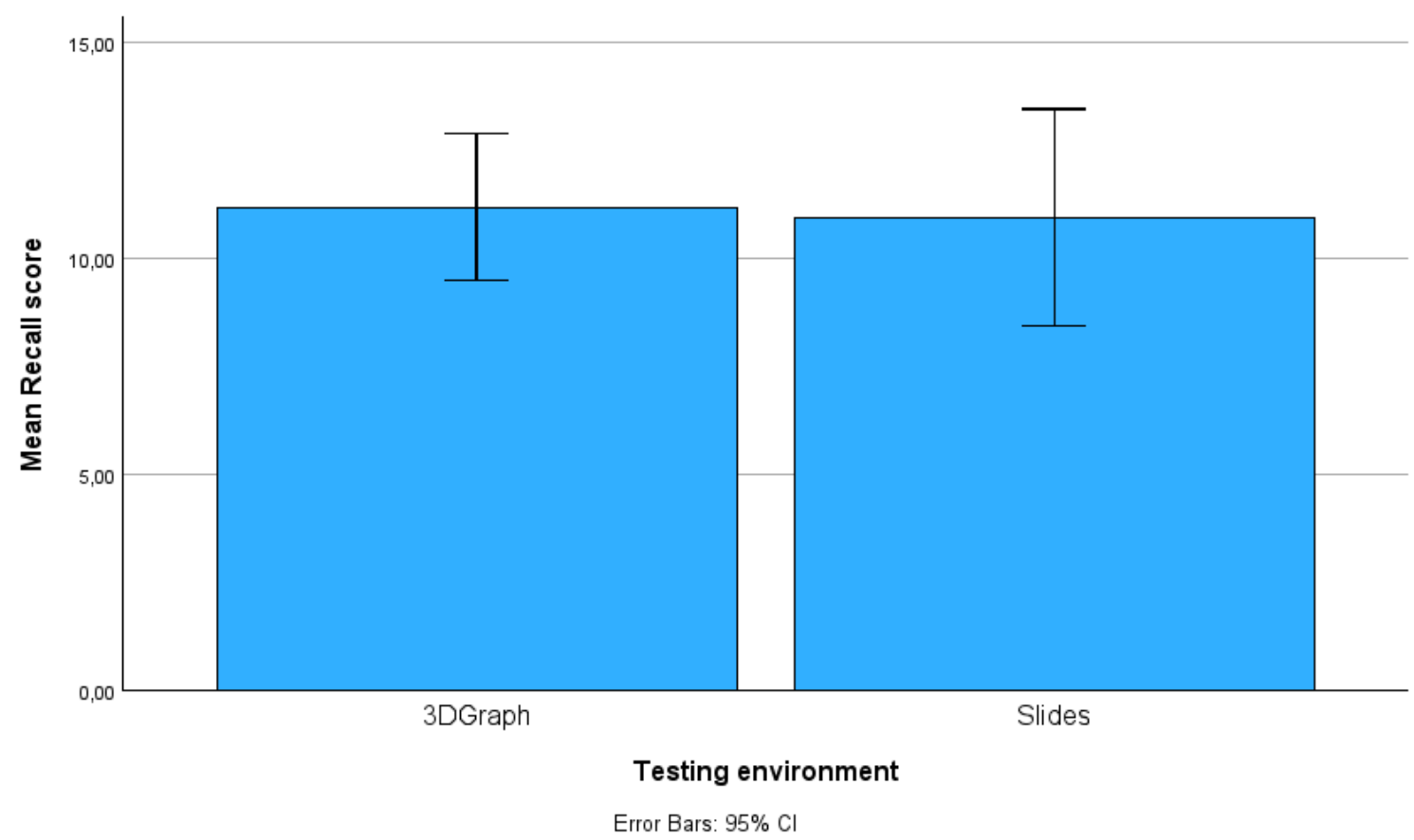

**Fig. 18.** Mean *recall score* by testing environment (the lower the better). Error bars show 95% CI

Fig. 18 shows that both groups had a similar recall score, meaning that participants did not remember specific linked data better with one visualisation condition. In this task, the user's ability to remember specific details after interacting with the testing environment was of interest, as short-term memory plays an important role in perceptually and conceptually integrating information usually dispersed across various models. When the number of models exceeds an engineer's span of control, the limited capacity of human working memory makes it practically impossible for them to retain all information from all models to reason successfully about the system. By gathering multiple viewpoints in a layered 3D graph representation, the intent is to minimise the importance of short-term memory, but this was not demonstrated in questions 4, 5, and 6. An issue might be that these questions evaluate how well the user retains information after interacting with a testing environment rather than the memorisation effort during the task. Thus, despite asking questions that evaluate tha ability to store and retrieve information a posteriori, future research should develop questions that require remembering information to be answered while performing the task.

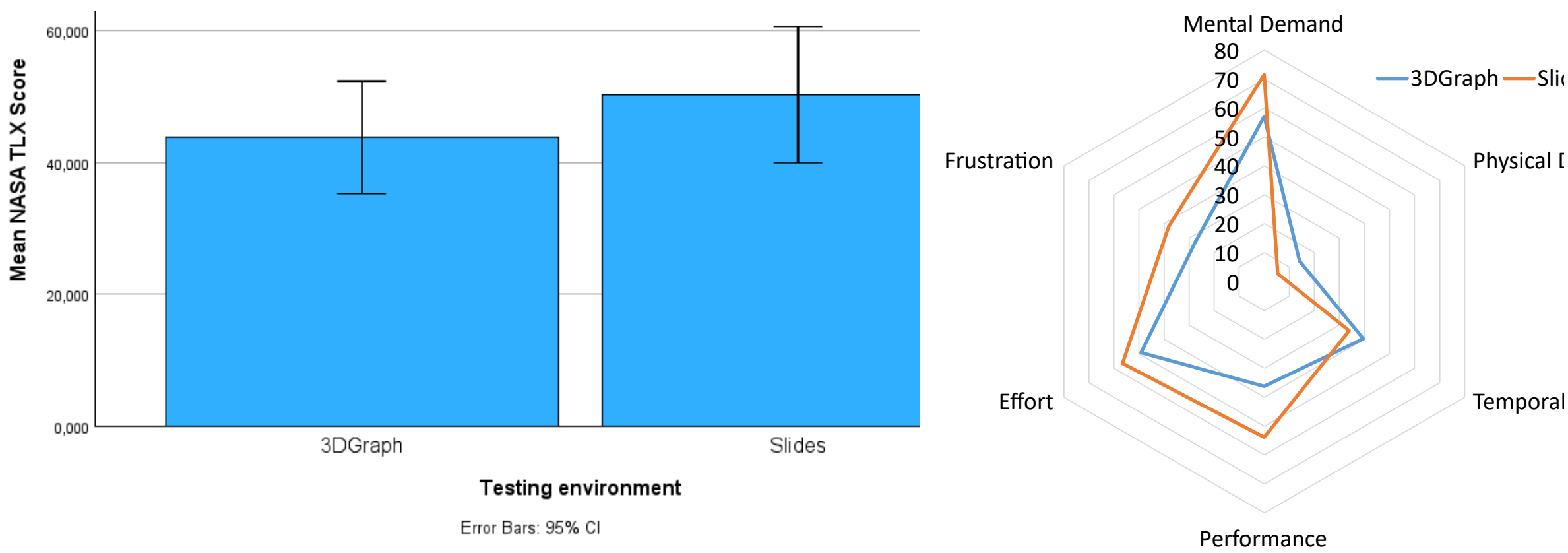

**Fig. 19.** Mean *NASA-TLX score* by testing environment (the lower the better) with 95% CI (left) and comparison (lower is better) of the perceived workload using the NASA-RTLX subscales (right)

Although the results are not statistically significant, Fig. 19 (left) indicates that participants with the Slides slightly perceived a greater workload than those with the 3D Graph. The experimenters instructed participants to complete the NASA-TLX questionnaire within 20 minutes, focusing on questions 1, 2, and 3, while using the testing environment. According to Fig. 19 (right), only two dimensions are more demanding. The physical demand is logically higher with the 3D Graph as a VR HMD requires participants to move their bodies to interact with the virtual environment.

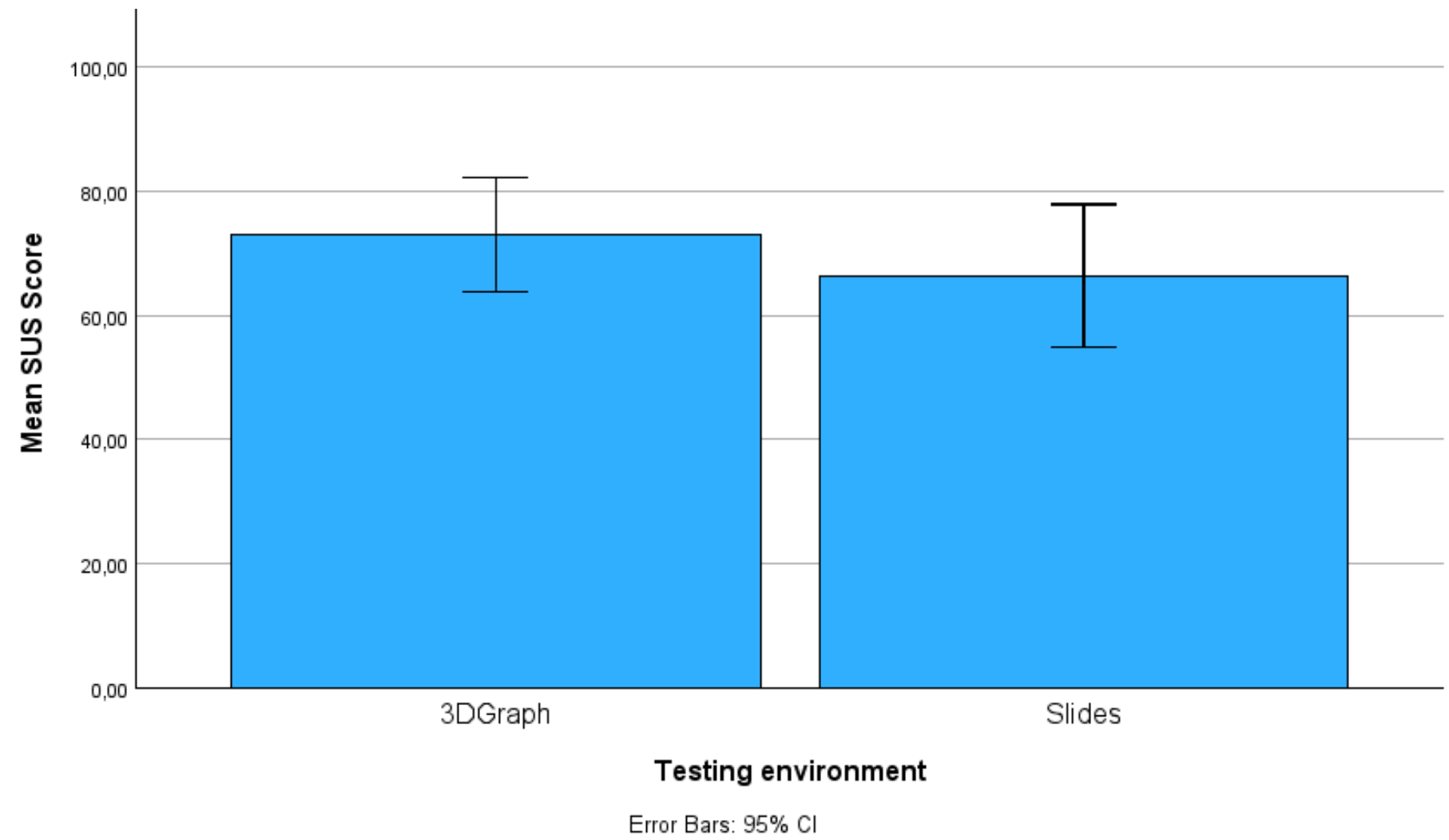

**Fig. 20.** Mean *system usability score* by testing environment (the higher the better). Error bars show 95% CI

Finally, in terms of perceived usability,
**Fig. 20** shows that the 3D Graph is more usable than the Slides. According to [80], an average score of 74 for the 3D Graph corresponds to "good usability" and falls into the "acceptable" category. On the other hand, a score of 66 for the Slides falls into the "OK" category, on the borderline between acceptable and unacceptable. In this case, the 3D Graph is clearly usable, while the slides are borderline. A more detailed analysis of the SUS topics (Fig. 21) indicates that participants agree that the 3D Graph is easy to use and would be willing to use it frequently.

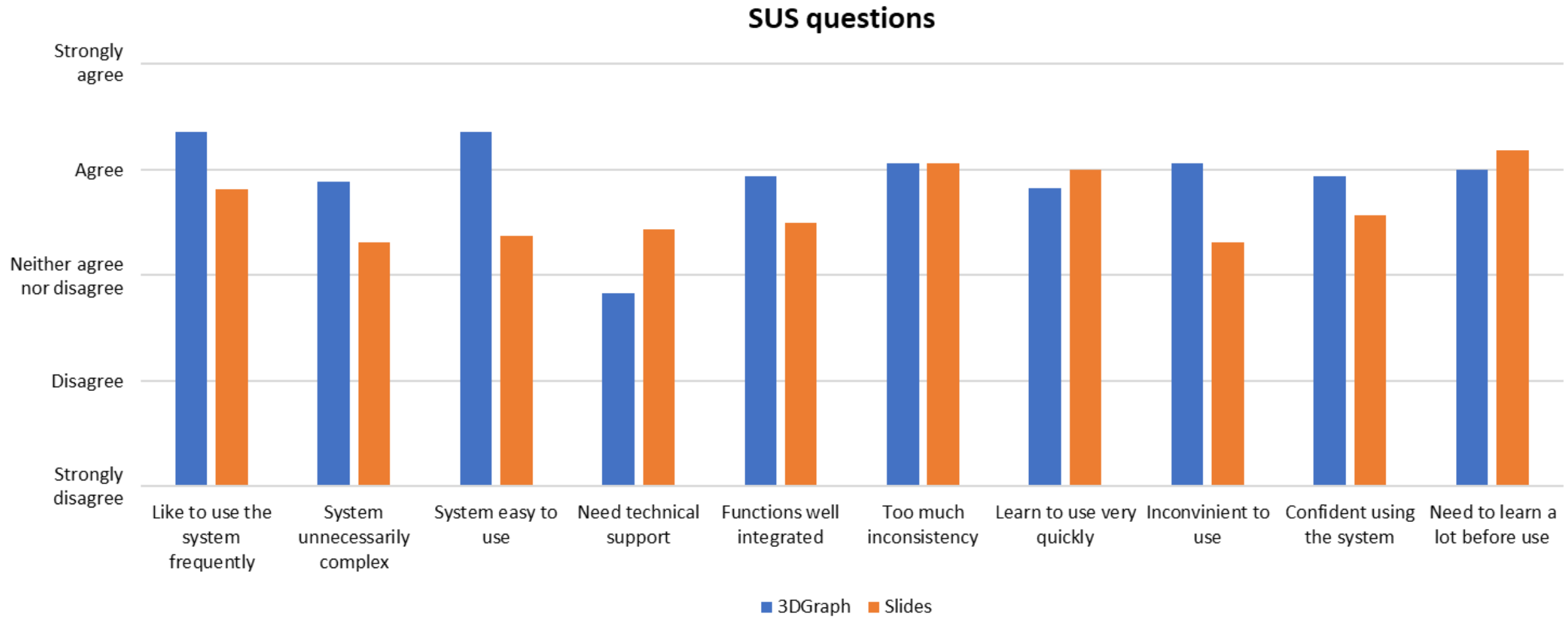

**Fig. 21.** Mean *system usability scale* topics by testing environment. Error bars show 95% CI

Regarding whether the experiment really measured what we wanted it to measure and whether our results really mean what we think and hope they mean, we previously discussed that the design of the experiment was not fully able to discriminate the test conditions. Nonetheless, this represents a preliminary effort to pave the way for more empirical research into the use of VR technologies in model-based design reviews. We are confident that the results are coherent as we meticulously analysed the data. Furthermore, although we anticipated that the VR environment would outperform the slides, it is not surprising to observe similar results due to the experimental design and the limited size of the dataset. Indeed, GraphXplore will eventually prove significantly superior to Slides when future experimental tasks require reviewing a large set of engineering linked data that cannot be processed by the human mind without advanced interactions and thoughtfully crafted immersive representations. To illustrate the added value of VR, the task should also necessitate analysing relatively long and complex paths. Finally, regarding reliability, if we repeat the experiment under the same conditions, we should obtain the same results, and others should as well. However, it would be more intriguing to enhance the design of the experiment as previously discussed.

Another threat to validity for this study's experimental design is the bundling of multiple independent variables, which inherently introduces potential confounding effects. The experimental setup compared two entirely distinct ecosystems: a 3D graph navigated via 6-DOF controllers in an immersive HMD, versus a 2D slide deck navigated via a keyboard and mouse on a standard desktop monitor. Consequently, it is difficult to definitively isolate whether the observed improvements in usability and reduced cognitive workload are primarily attributable to the representation format (the layered 3D graph itself), the visualisation device (the immersive stereoscopic HMD), or the interaction mechanism (natural 3D spatial manipulation). While these confounding variables could have been avoided by using the same interaction and visualisation devices for both groups (e.g., testing the 3D graph on a standard 2D desktop monitor), doing so would fail to represent how these immersive MBSE tools are actually intended to be used in practice, thereby compromising the ecological validity of the study. Nevertheless, future experimental designs must attempt to systematically decouple these variables to pinpoint the precise source of the cognitive benefits.

# 7. Conclusion

As the use of integrated digital models to engineer technical systems increases, this paper has explored the potential of VR technology to examine a linked set of digital artefacts throughout a product's life cycle. This study, situated at the intersection of Human-Computer Interaction and MBSE, aimed to enhance our understanding of the selection of representations for encoding MBSE data, with a particular focus on digital threads. The existing literature presents only early and limited prototypes for integrating MBSE and VR, lacking empirical evidence, and no research has explored the usefulness of examining MBSE digital threads in VR. Thus, this paper sought to enhance the perceptual and conceptual integration processes by linking all relevant visual elements from different perspectives in an immersive and interactive layered 3D graph representation, offering a holistic system view.

Because this baseline study focused on fundamental information retrieval tasks, it is crucial to explicitly differentiate between the study's validated conclusions and its exploratory findings:

- **Validated Conclusions:** The experimental data statistically validates that the immersive layered 3D graph (GraphXplore) provides significant, measurable benefits in terms of perceived usability (SUS) and reduced cognitive workload (NASA-TLX) compared to a traditional slide-based setup during foundational MBSE design reviews.
- **Exploratory Findings:** Conversely, the findings regarding task completion time, recall scores, and accuracy remain strictly exploratory. Because the evaluation relied on simple retrieval tasks within a small-scale dataset, the comparable performance between the two environments cannot be definitively interpreted as a lack of VR efficacy. Rather, these exploratory null results suggest that traditional 2D slides remain highly efficient for small-scale lookups, highlighting the necessity for future experiments utilising large, enterprise-scale digital threads where the cognitive limits of 2D navigation are truly exceeded.

After introducing GraphXplore, we conducted a comparative study with the traditional on-screen setup, where a handcrafted PowerPoint presentation contains screenshots of models reviewed using a desktop PC, keyboard, and mouse. The experiment focused on understanding an MBSE design, as comprehending models is the most fundamental task in design review. Our findings suggest that the accuracy of answers, completion time, recall scores, and perceived confidence in both environments are comparable. However, while not statistically significant, usability was rated higher and workload lower when using the virtual environment.

Therefore, based on the Design Science Validity Framework [81], this paper provides several types of contributions to the existing MBSE body of knowledge. First, criterion-related efficacy validity was demonstrated, showing that the design artefact – the immersive layered 3D graph – provides benefits in terms of usability and cognitive workload. Second, causal efficacy validity supports causal claims through evaluation by comparing the efficacy of the focal artefact to that of a handcrafted PowerPoint presentation. The third contribution includes the development of a functional virtual environment, a benchmark exercise, and an evaluation protocol. Although there are various ways to enhance the demonstrator and design experiments, this study is unique in exploring digital threads and is among the few in MBSE that offers empirical data, supporting evidence-based research.

Building upon these findings, the key insights derived from this experiment are twofold. First, immersive 3D spatial navigation seems to inherently offload the cognitive burden required to mentally integrate fragmented MBSE views, making it easier for users to conceptualise the system as a whole. Second, establishing a high baseline usability for GraphXplore on foundational tasks demonstrates that engineers can comfortably adopt and navigate an immersive 3D graph visual metaphor, without being hindered by the steep learning curves typically associated with new MBSE interfaces. Consequently, these insights drive several critical implications for future research:

- **Scaling to Enterprise Complexity:** Future experiments must evaluate GraphXplore using large, densely connected industrial digital threads to fully quantify the tipping point where traditional 2D representations fail and immersive 3D graphs become strictly necessary.
- **AI-Assisted Exploration:** Future iterations of the system will integrate conversational generative AI agents, allowing users to query and filter massive architectural models using natural language rather than manual navigation alone [82].
- **Collaborative Design Reviews:** Future research must transition from single-user evaluations to multi-user collaborative environments. Today, we have developed asymmetric multi-device collaboration capabilities across HMDs, PCs, Powerwalls, and CAVEs in a completely new solution[3], but evaluation is ongoing. This allows a multidisciplinary team to review digital threads and designs or co-design a solution while considering the concerns of various stakeholders, each using their own visual external device representations. New experimental protocols will be designed to measure how the collaborative virtual environment influences the collaboration and cognitive processes of systems architects and non-expert stakeholders during concurrent engineering activities.

**Data availability** Data, including the SPSS statistical analysis file, is available upon request to the corresponding author.

# Declarations

**Competing interests** The authors declare they have no competing interests.

[3] https://www.youtube.com/watch?v=c5bjCI4g2jQ